\documentclass[pmlr]{jmlr}

\RequirePackage{graphicx}
 \usepackage{booktabs}
 \usepackage{enumitem}
\usepackage{longtable}
\usepackage{pdflscape}
\makeatletter
\def\set@curr@file#1{\def\@curr@file{#1}} 
\makeatother
\usepackage[load-configurations=version-1]{siunitx} 

\usepackage{subcaption, graphicx}
\usepackage{amsmath, amsfonts, amssymb}

\DeclareMathOperator*{\argmax}{arg\,max}

\newtheorem{defn}{Definition}

\theorembodyfont{\upshape}
\theoremheaderfont{\scshape}
\theorempostheader{:}
\theoremsep{\newline}

\jmlrvolume{}
\jmlryear{2025}
\jmlrworkshop{}

\title[Public Perception of Fairness in Regression-based Models]{Fairness Perceptions in Regression-based Predictive Models}

\author{\Name{Mukund Telukunta}
       \Email{mt3qb@mst.edu}\\ 
       \addr Computer Science\\
       Missouri University of Science and Technology\\
       Rolla, Missouri, USA
       \AND
       \Name{Venkata Sriram Siddhardh Nadendla}
       \Email{nadendla@mst.edu}\\ 
       \addr Computer Science\\
       Missouri University of Science and Technology\\
       Rolla, Missouri, USA
       \AND
       \Name{Morgan Stuart}
       \Email{Morgan.Stuart@unos.org}\\ 
       \addr United Network for Organ Sharing (UNOS)\\
       Virginia, USA
       \AND
       \Name{Casey Canfield}
       \Email{canfieldci@mst.edu}\\ 
       \addr Engineering Management and Systems Engineering\\
       Missouri University of Science and Technology\\
       Rolla, Missouri, USA} 

\begin{document}

\maketitle

\begin{abstract}
Regression-based predictive analytics used in modern kidney transplantation is known to inherit biases from training data. This leads to social discrimination and inefficient organ utilization, particularly in the context of a few social groups. Despite this concern, there is limited research on fairness in regression and its impact on organ utilization and placement. This paper introduces three novel divergence-based group fairness notions: ($i$) independence, ($ii$) separation, and ($iii$) sufficiency to assess the fairness of regression-based analytics tools. In addition, fairness preferences are investigated from crowd feedback, in order to identify a socially accepted group fairness criterion for evaluating these tools. A total of 85 participants were recruited from the Prolific crowdsourcing platform, and a Mixed-Logit discrete choice model was used to model fairness feedback and estimate social fairness preferences. The findings clearly depict a strong preference towards the separation and sufficiency fairness notions, and that the predictive analytics is deemed fair with respect to gender and race groups, but unfair in terms of age groups.
\end{abstract}

\section{Introduction}
The United Network for Organ Sharing (UNOS) introduced a regression-based predictive analytics tool \citep{mcculloh2023experiment} to assist transplant professionals in decision-making at the time of an organ offer. The main objective of UNOS' predictive analytics tool (UPAT) is to address the critical challenge of allocating limited kidneys to waitlisted recipients, amidst high demand.
However, disparities in kidney transplantation practices persist, although they are mostly driven by medical and biological factors. For example, African-Americans exhibit a more diverse human leukocyte antigen (HLA) profile, complicating the search for compatible deceased donor kidneys. Only 0.7\% of Black patients receive fully HLA-A/B/DR/DQ matched transplants, compared to 8.1\% of White patients, leading to significantly longer wait times and lower transplant rates \citep{bekbolsynov2022improving}. Women, often sensitized by pregnancy, develop elevated panel reactive antibody (PRA) levels, reducing their compatibility with deceased donor organs and contributing to lower transplantation rates \citep{salas2022sex}. Additionally, women are more likely to be misclassified as frail due to biases in assessment methods; non-frail women on hemodialysis are more frequently labeled frail compared to non-frail men, decreasing their chances of waitlisting \citep{harding2023sex}. Older patients ($\geq$ 65), face higher rates of comorbidities such as diabetes, which increase the risk of post-transplant complications like infection and rejection \citep{veroux2012age}. As a result, transplant centers often prioritize patients with lower-risk profiles in order to maximize the number of life years gained from each transplant.

Although these disparities are medically justified, predictive models trained on such data may inadvertently perpetuate further inequities, leading to unjustifiable discrimination. If these biases are not monitored, research shows that informational interventions can worsen inequality \citep{veinot2018good}. Consequently, there is an urgent need to evaluate UPAT from an algorithmic fairness perspective. However, the majority of the group fairness literature has focused on classification problems, with relatively few contributions to fairness in regression \citep{agarwal2019fair, chzhen2020fair}. Notably, these proposed notions in regression mainly focus on comparing loss between two groups, which makes them less interpretable in practical applications such as kidney transplantation. Hence, inspired by prior work \citep{barocas2019fairness, steinberg2020fairness}, this paper proposes three novel group fairness notions for regression: \emph{independence}, \emph{separation} and \emph{sufficiency}. These notions use Kullback-Leibler (KL) divergence to evaluate the similarity between outcome distributions (predicted or actual) across different social groups. Note that inherent tradeoffs may exist amongst group fairness notions as it could be impossible to satisfy multiple fairness criteria simultaneously \citep{chouldechova2017fair, kleinberg2016inherent}. Consequently, practitioners must carefully evaluate which fairness criterion is best suitable for a given system.

Therefore, this paper examines public perceptions of fairness in UPAT, along with identifying the most preferred fairness notion for its evaluation. A human-subject survey experiment was conducted with 85 participants recruited using Prolific crowd-sourcing platform to assess the fairness of the UPAT. Their responses were modeled using a mixed-logit framework from discrete choice theory, assuming each participant exhibits a weighted preference over the proposed fairness notions. The study estimates social fairness preferences by employing a projected gradient-based learning algorithm that minimizes social feedback regret. While previous research has explored human perceptions of fairness \citep{srivastava2019mathematical, harrison2020empirical, lavanchy2023applicants, telukunta2024fairpreference}, it has predominantly focused on classification tasks. To the best of our knowledge, this paper presents the very first attempt to explore human perceptions of fairness in a regression setting. In addition to the survey experiment, the proposed preference learning algorithm is evaluated using simulated participant feedback. The results demonstrate a quick convergence of the feedback regret within 50 training epochs. Our experimental findings suggest that the recruited participants prefer the separation and sufficiency notions and judge the UPAT tool to be completely fair with respect to gender and race, while unfair with respect to age.

\subsection*{Generalizable Insights about Machine Learning in the Context of Healthcare}
This group fairness criteria and the preference learning algorithm proposed to evaluate a regression-based decision-support system can also be adopted to investigate fairness of AI models in other healthcare applications such as liver transplantation \citep{gotlieb2022promise} and cancer risk prediction \citep{shipe2019developing}. In addition,  the fairness feedback elicitation protocol can be generalized to any potential stakeholder in the kidney transplant pipeline.

\section{Experiment Design}
The objective of the survey experiment is to collect feedback from public participants regarding the fairness of the UPAT \citep{mcculloh2023experiment} tool. UPAT was deployed through the DonorNet\textsuperscript{\textcopyright} platform which informs transplant surgeons about the quality of the incoming kidney offer and the prospect of the recipient's health if the current offer is refused. This enables decision-makers (e.g. transplant surgeons) to make reliable decisions on whether to accept or refuse a kidney offer. Specifically, the UPAT provides two different decision-support predictions:
\begin{enumerate}[leftmargin=*]
\item \textbf{Time-To-Next-Offer (TTNO)}: The predicted length of time the recipient could wait for another organ offer. These predictions are based on Log-Logistic Accelerated Failure Time model with seven different characteristics (e.g. blood type, Estimated Post-Transplant Survival (EPTS)) of the candidate. The model estimates TTNO for both Kidney Donor Profile Index (KDPI) $<$30 and KDPI $<$50, where KDPI is a numerical score ranging from 0 to 100 that quantifies the quality of a deceased donor kidney, with lower values indicating longer predicted graft survival once transplanted. However, in this paper, the predictions pertaining to KDPI$<$30 are only considered. Note that the TTNO predictions are independent of any demographic membership of the recipient.
\item \textbf{Mortality Likelihood before TTNO}: The candidate's predicted survival likelihood without transplant before TTNO. These predictions are based on a Cox Proportional Hazards model optimized to estimate the time until removal for death or too sick to transplant, with 10 different characteristics (e.g. diabetes status, age, body mass index (BMI)) of the candidate as input features. Note that age attribute is the only demographic feature included in this model.
\end{enumerate}
Note that the mortality likelihood model does not know the predicted TTNO, nor vice versa, during execution. These models were trained on recipient dataset from the year 2019 onwards. A group of 15 transplant programs piloted the Predictive Analytics tool in 2022, and saw their acceptance rate increase from 16.8\% to 19.7\% on offers viewed through DonorNet \citep{mcculloh2023experiment}. The UPAT was deployed in January 2023 and is currently used by transplant surgeons in the United States.

\subsection{Datasets and Preprocessing} 
The participants are asked to rate the fairness of TTNO and mortality likelihood predictions from the UPAT for various kidney matching instances spanning years 2021 and 2022. These kidney matching instances are obtained from the Standard Transplant Analysis and Research (STAR) datasets, requested from the Organ Procurement and Transplant Network (OPTN). The STAR datasets consists of anonymized data at the patient level, covering transplant recipients, donors, and their matches since 1987. The following preprocessing steps were performed to construct the data required for the survey. First, recipients younger than 17 were omitted due to the distinct challenges associated with pediatric transplantation \citep{magee2004pediatric}. The STAR dataset typically records multiple instances where a deceased donor kidney is matched with thousands of potential recipients. Since it is impractical for any participant to scrutinize every possible match, the list of potential recipients per deceased donor was limited to 10. This selection always includes at least one recipient (at most two) who actually received the kidney for transplantation. The remaining 9 (or 8, if two transplantations were performed) recipients were chosen to closely match the demographics (age, race and gender) of all potential recipients for that donor. The preprocessed dataset included data of 6 deceased donor profiles for the year 2021 and 4 for 2022. From this processed dataset, 10 data-tuples (each consisting of one donor and 10 recipients) were selected. In total, 100 random potential recipients were chosen from the preprocessed dataset for our survey experiment, ensuring a representation of diversity across various social groups similar to that seen in the complete STAR files for 2021 and 2022 (shown in Table \ref{Tab: Demographic Comparison}). The UPAT was then applied to this sample to obtain both TTNO and mortality rate predictions for every potential recipient within each deceased donor.

\subsection{Participant Survey Design}

\begin{table}[!t]
\begin{minipage}{.55\linewidth}
\caption{Demographic Comparison}
\centering
\begin{tabular}{ c  c  c }
\toprule
& \multicolumn{2}{c}{\textbf{\% Data Points}} 
\\
& Sampled & STAR
\\
& Data-Tuples & 2021+2022
\\\bottomrule
\\[-2ex]
Male & 62\% & 64\%
\\
Female & 38\% & 35\%
\\[1ex]
White & 46\% & 41\%
\\
Black & 25\% & 28\%
\\
Hispanic & 19\% & 21\%
\\
Asian & 8\% & 7\%
\\
Other & 2\% & 3\%
\\[1ex]
$<=$50 years & 50\% & 54\%
\\
$>$50 years & 50\% & 46\%
\\ \bottomrule
\end{tabular}
\label{Tab: Demographic Comparison}
\end{minipage} 
\begin{minipage}{.4\linewidth}
\caption{An Example of Donor Characteristics}
\centering
\begin{tabular}{l l}
\hline
\multicolumn{2}{c}{\textbf{DONOR}}
\\ \hline
\textbf{Age} & 30
\\
\textbf{Race} & Black
\\
\textbf{Gender} & Female
\\
\textbf{Kidney Quality} & 45
\\\hline
\end{tabular}
\label{Tab: DD Characteristics}
\end{minipage}%
\end{table}

\begin{figure*}[!t]
\centering
\includegraphics[width=\textwidth]{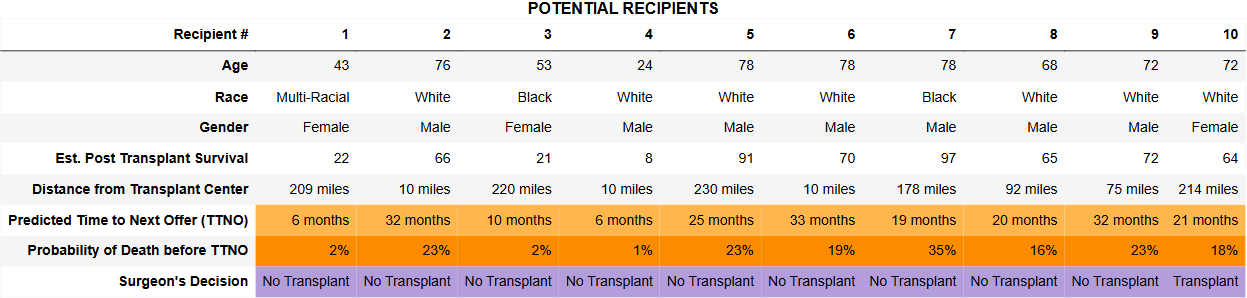}
\caption{An Example of Recipient Characteristics}
\label{Fig: Recipient Characteristics}
\end{figure*}

\begin{figure*}[!t]
\centering
\includegraphics[width=\textwidth]{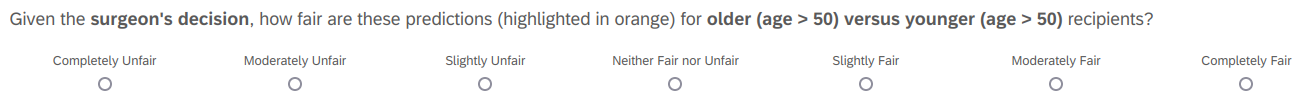}
\includegraphics[width=\textwidth]{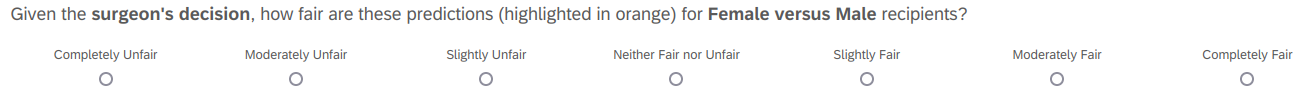}
\includegraphics[width=\textwidth]{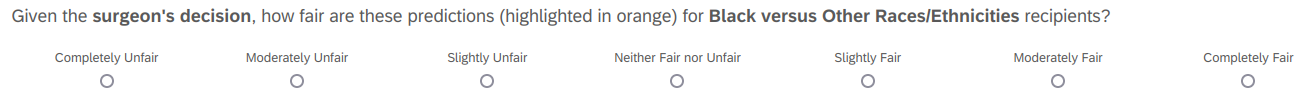}
\caption{Three Questions Presented to the Participants for Each Data Tuple}
\label{Fig: Survey Questions}
\end{figure*}

In this experiment, each participant is presented with 10 data-tuples, each consisting of two data tables. The first table includes deceased donor characteristics, such as age, race, gender, and KDPI score, as shown in the Table \ref{Tab: DD Characteristics}. The second table presents characteristics of 10 recipients matched with this donor. This includes each recipient's age, race, gender, EPTS score, distance from the transplant center, TTNO and mortality rate predictions from the UPAT, and the surgeon's decision (transplant or no transplant), as depicted in Figure \ref{Fig: Recipient Characteristics}. Participants were asked to rate the fairness of the UPAT using a Likert scale ranging from 1 to 7, where 1 indicates complete unfairness and 7 denotes complete fairness. Specifically, participants were prompted to assess the fairness of the UPAT in the context of (i) older recipients (age$>=$50) versus younger recipients (age$<$50), (ii) female versus male recipients, and (iii) Black recipients versus recipients from other racial backgrounds, as depicted in Figure \ref{Fig: Survey Questions}. Please refer Appendix \ref{App: Complete Survey Design} for more details about the survey design.


\subsection{Participants Demographics}
The survey experiment, conducted on Prolific in December 2023,
initially recruited 85 participants for the study. After excluding 2 participants who did not pass the attention check questions, 83 participants were selected. On average, participants spent 20 minutes completing the survey and received compensation at a rate of \$11.58 per hour. Table \ref{Tab: participant demographics} presents the demographic breakdown of the participants. The demographic composition included a lower percentage of Hispanic participants (3.4\%), a higher percentage of Black participants (19\%), a more educated group (51\%), and a younger cohort (65\%) compared to the 2021 U.S. Census \citep{Census2021ACS}. Such demographic discrepancies are common when using crowd-sourcing platforms for participant recruitment. Recruiting older participants can be challenging due to reduced social media engagement \citep{turner2020recruiting}. Moreover, Prolific is known for its tendency to attract younger participants \citep{charalambidesProlific}.

\begin{table}[!t]
\caption{Participants Demographics Compared to 2021 U.S. Census Data.}
\centering
\begin{tabular}{l r r}
\hline
\textbf{Demographic Attribute} & \textbf{Our Survey} & \textbf{Census}
\\ \hline
18-25 & 8\% & 13\%
\\[-0.5ex]
25-40 & 57\% & 26\%
\\[-0.5ex]
40-60 & 29\% & 32\%
\\[-0.5ex]
$>$60 & 6\% & 22\%
\\ \hline
White & 60\% & 59\%
\\[-0.5ex]
Black & 19\% & 12\%
\\[-0.5ex]
Asian & 12\% & 5.6\% 
\\[-0.5ex]
Hispanic & 3.4\% & 18\%
\\[-0.5ex]
Other & 5.6\% & 9\%
\\ \hline
Male & 49\% & 49.5\%
\\[-0.5ex]
Female & 49\% & 50.5\%
\\[-0.5ex]
Non-binary & 2\% & -
\\ \hline
High School or equivalent & 18\% & 26.5\%
\\[-0.5ex]
Bachelor's (4 year) & 40\% & 20\%
\\[-0.5ex]
Associate (2 year) & 15\% & 8.7\%
\\[-0.5ex]
Some college & 12\% & 20\%
\\[-0.5ex]
Master's & 11\% & 13\%
\end{tabular}
\label{Tab: participant demographics}
\end{table}

\section{Method}
Let $x \in \mathcal{X}$ denote the donor-recipient input characteristics comprising clinical and demographic features such as age, race and blood type. Consider $g(\cdot)$ as the UPAT and $\hat{y} = g(x) \in \mathcal{Y}$ denote the set of predictions $\hat{y} = [\hat{y}_T, \hat{y}_D]$, where $\hat{y}_T, \hat{y}_D$ indicate the time-to-next-offer and the mortality rate predictions respectively. Let $z \in \{0, 1\}$ denote the binary decision made by the transplant surgeon, where $z=1$ indicates organ offer acceptance and $z=0$ indicates refusal. Furthermore, let $\mathcal{X}_m, \mathcal{X}_{m'} \in \mathcal{X}$ denote advantaged and disadvantaged social groups (e.g. age, race), respectively. 

\subsection{Divergence-Based Group Fairness Notions \label{Sec: Divergence Measures}} 
Various group fairness notions have been proposed in classification tasks, each aiming to evaluate and compare statistical measures across different social groups \citep{MoritzOpportunities, chouldechova2018frontiers}. These notions can be broadly categorized into three primary frameworks \citep{barocas2019fairness}: (i) \emph{independence}, which ensures that predictions are statistically independent of social group membership $P(\hat{y} \ | \ \mathcal{X}_m) = P(\hat{y})$; (ii) \emph{separation}, which requires that predictions remain independent of social groups when conditioned on the actual outcome $P(\hat{y} \ | \ z, \mathcal{X}_m) = P(\hat{y} \ | \ z)$; and (iii) \emph{sufficiency}, which ensures that the likelihood of the actual outcome is consistent across social groups when conditioned on the predictions $P(z \ | \ \hat{y}, \mathcal{X}_m) = P(z \ | \ \hat{y})$. This paper extends these fairness frameworks to the regression setting by comparing the probability distributions of predictions (or surgeon decisions) across two social groups, $\mathcal{X}_m, \mathcal{X}_{m'} \in \mathcal{X}$. Specifically, the divergence between the probability distributions of two groups is quantified using the Kullback-Leibler (KL) divergence, denoted as $\mathbb{D}_{KL}(\cdot ||\cdot)$. Formally, this paper proposes three novel fairness notions defined as follows.

\begin{defn}[Independence]
A decision-support system is said to satisfy independence if
\begin{equation}
\mathbb{D}_{KL}\Big(P(\hat{y} \ | \ x \in \mathcal{X}_m) \ \Big|\Big| \ P(\hat{y} \ | \ x \in \mathcal{X}_{m'})\Big) = 0,
\label{Eqn: Independence}
\end{equation}
for any two socials groups $\mathcal{X}_m, \mathcal{X}_{m'} \in \mathcal{X}$.
\label{Defn: Independence}
\end{defn}
In other words, independence requires that the prediction distributions $\pi(\hat{y}) = P(\hat{y} \ | \ x \in \mathcal{X}_m)$ and $\pi'(\hat{y}) = P(\hat{y} \ | \ x \in \mathcal{X}_{m'})$ be similar (i.e., zero divergence) for different groups, ensuring predictions are not influenced by demographic attributes. In the case of UPAT, the predictions $\hat{y}_T$ and $\hat{y}_D$ are independent with each other. Then, Equation \eqref{Eqn: Independence} can be written as 
\begin{equation}
\begin{array}{lcl}
\mathbb{D}_{KL}\Big(\pi(\hat{y}) \ \Big|\Big| \ \pi'(\hat{y})\Big) &=& \displaystyle \iint \pi(\hat{y}_T)\cdot \pi(\hat{y}_D) \cdot log \left( \frac{\pi(\hat{y}_T)\cdot \pi(\hat{y}_D)}{\pi'(\hat{y}_T)\cdot \pi'(\hat{y}_D)} \right) d\hat{y}_T \ d\hat{y}_D
\\[3ex]
&=& \displaystyle \int_0^{\infty} \pi(\hat{y}_T)\cdot log \left( \frac{\pi(\hat{y}_T)}{\pi'(\hat{y}_T)} \right) d\hat{y}_T + \displaystyle \int_0^1 \pi(\hat{y}_D) \cdot log \left( \frac{\pi(\hat{y}_D)}{\pi'(\hat{y}_D)} \right) d\hat{y}_D
\\[3ex]
&=& \mathbb{D}_{KL}\left(\pi(\hat{y}_T) \ || \ \pi'(\hat{y}_T)\right) + \mathbb{D}_{KL}\left(\pi(\hat{y}_D) \ || \ \pi'(\hat{y}_D)\right)
\end{array}
\label{Eqn: Independence for y_T and y_D}
\end{equation}
This indicates that the UPAT is said to satisfy independence if the distribution of time-to-next-offer predictions as well as the distribution of mortality rate predictions are same across any two social groups. Note that the predictions $\hat{y}_T, \hat{y}_D$ follow Weibull distribution. Therefore, the $\mathbb{D}_{KL}(\cdot || \cdot)$ terms in Equation \eqref{Eqn: Independence for y_T and y_D} can be solved using the formulation from \cite{bauckhage2014kernel}, as follows.
\begin{equation}
\begin{array}{lcl}
\mathbb{D}_{KL}\left(\pi(\hat{y}_T) \ || \ \pi'(\hat{y}_T)\right) &=& \displaystyle \log \left( \frac{k_T}{\lambda_T^{k_T}} \right)
- \log \left( \frac{k'_T}{(\lambda'_T)^{k'_T}} \right) + \ \displaystyle (k_T - k'_T)\left[\log\lambda_T - \frac{\gamma}{k_T}\right]
\\[3.5ex]
&& \qquad + \displaystyle \left( \frac{\lambda_T}{\lambda'_T} \right)^{k'_T} \Gamma \left( \frac{k'_T}{k_T} + 1 \right) - 1,
\end{array}
\label{Eqn: KLD Weibull y_T}
\end{equation}
where $k_T$ and $\lambda_T$ represent the shape and scale parameters of the distribution $\pi(\hat{y}_T)$, respectively, and $k'_T$ and $\lambda'_T$ represent the shape and scale parameters of the distribution $\pi'(\hat{y}_T)$, respectively. Here, $\gamma \approx 0.5772$ is the Euler-Mascheroni constant and $\Gamma(\cdot)$ denotes the Gamma function. Similarly,
\begin{equation}
\begin{array}{lcl}
\mathbb{D}_{KL}\left(\pi(\hat{y}_D) \ || \ \pi'(\hat{y}_D)\right) &=& \displaystyle \log \left( \frac{k_D}{\lambda_D^{k_D}} \right) 
- \log \left( \frac{k'_D}{(\lambda'_D)^{k'_D}} \right) + \displaystyle (k_D - k'_D)\left[\log\lambda_D - \frac{\gamma}{k_D}\right]
\\[3ex]
&& \qquad + \displaystyle \left( \frac{\lambda_D}{\lambda'_D} \right)^{k'_D} \Gamma \left( \frac{k'_D}{k_D} + 1 \right) - 1.
\end{array}
\label{Eqn: KLD Weibull y_D}
\end{equation}

\begin{table}[!t]
\caption{Proposed Group Fairness Notions for Regression-setting}
\centering
\begin{tabular}{c l l}
\toprule
Index ($\ell$) & \textbf{Group Fairness Notion} & \textbf{Fairness Score} $\phi_{\ell}$  \\
\bottomrule
\\[-2ex]
1 & Independence & $\phi_{1} = \mathbb{D}_{KL}\Big(P(\hat{y} \ | \ x \in \mathcal{X}_m) \ \Big|\Big| \ P(\hat{y} \ | \ x \in \mathcal{X}_{m'})\Big)$
\\[1.5ex]
2 & Separation & $\phi_{2} = \mathbb{D}_{KL}\Big(P(\hat{y} \ | \ z, x \in \mathcal{X}_m) \ \Big|\Big| \ P(\hat{y} \ | \ z, x \in \mathcal{X}_{m'})\Big)$
\\[1.5ex]
3 & Sufficiency & $\phi_{3} = \mathbb{D}_{KL}\Big(P(z \ | \ \hat{y}, x \in \mathcal{X}_m) \ \Big|\Big| \ P(z \ | \ \hat{y}, x \in \mathcal{X}_{m'})\Big)$
\\\bottomrule
\end{tabular}
\label{tab:my_label}
\end{table}

\begin{defn}[Separation]
A decision-support system is said to satisfy separation if
\begin{equation}
\mathbb{D}_{KL}\Big(P(\hat{y} \ | \ z, x \in \mathcal{X}_m) \ \Big|\Big| \ P(\hat{y} \ | \ z, x \in \mathcal{X}_{m'})\Big) = 0,
\label{Eqn: Separation}
\end{equation}
for any two socials groups $\mathcal{X}_m, \mathcal{X}_{m'} \in \mathcal{X}$.
\label{Defn: Separation}
\end{defn}
Separation ensures that the prediction distributions $\pi_z(\hat{y}) = p(\hat{y} \ | \ z, x \in \mathcal{X}_m)$ and $\pi'_z(\hat{y}) = p(\hat{y} \ | \ z, x \in \mathcal{X}_{m'})$ conditioned with the surgeon's decision $z$ be similar for the social groups $\mathcal{X}_m$ and $\mathcal{X}_{m'}$, respectively. Furthermore, given that the predictions $\hat{y}_T, \hat{y}_D$ are independent and the surgeon's decision $z \in \{0, 1\}$, Equation \eqref{Eqn: Separation} can be written as
\begin{equation}
\begin{array}{l}
\mathbb{D}_{KL}\Big(\pi_z(\hat{y}) \ \Big|\Big| \ \pi'_z(\hat{y})\Big)
\\[2ex]
\qquad = \displaystyle \sum_{z \in [0, 1]} p(\mathcal{Z} = z) \cdot \Big( \mathbb{D}_{KL}\left(\pi_{\mathcal{Z} = z}(\hat{y}_T) \ || \ \pi'_{\mathcal{Z} = z}(\hat{y}_T)\right) + \mathbb{D}_{KL}\left(\pi_{\mathcal{Z} = z}(\hat{y}_D) \ || \ \pi'_{\mathcal{Z} = z}(\hat{y}_D)\right) \Big).
\end{array}
\label{Eqn: Separation for y_t, y_d and z}
\end{equation}
Since $\hat{y}_T, \hat{y}_D$ follow Weibull distribution, the $\mathbb{D}_{KL}(\cdot || \cdot)$ terms in the above equation can be solved using the Equations \eqref{Eqn: KLD Weibull y_T} and \eqref{Eqn: KLD Weibull y_D}. This formulation helps in understanding how the system's predictions vary with \emph{expert-human} (i.e., surgeon) decisions and across different groups, revealing potential biases. For example, assume that the UPAT predicts a short waiting time (say 3 months) and a low mortality rate (e.g., 5\%) for a recipient. Given this information, the transplant surgeon decides not to transplant immediately, indicating that the UPAT predictions are relative to the surgeon's decisions. On the other hand, UPAT consistently predicts long waiting times, such as 30 months, and a high probability of death, like 60\%, for a particular group. If the surgeon decides not to transplant immediately, even though the system indicates high risk, this may suggest that the system is deviating.

\begin{defn}[Sufficiency]
A decision-support system is said to satisfy sufficiency if
\begin{equation}
\mathbb{D}_{KL}\Big(P(z \ | \ \hat{y}, x \in \mathcal{X}_m) \ \Big|\Big| \ P(z \ | \ \hat{y}, x \in \mathcal{X}_{m'})\Big) = 0,
\label{Eqn: Sufficiency}
\end{equation}
for any two socials groups $\mathcal{X}_m, \mathcal{X}_{m'} \in \mathcal{X}$.
\label{Defn: Sufficiency}
\end{defn}
The terms $\pi_{\hat{y}}(z) = P(z \ | \ \hat{y}, x \in \mathcal{X}_m)$ and $\pi'_{\hat{y}}(z) = P(z \ | \ \hat{y}, x \in \mathcal{X}_{m'})$ denote the likelihood of the surgeon's decision $z$ given the predictions $\hat{y} = [\hat{y}_T, \hat{y}_D]$. In other words, sufficiency ensures that the predictions are equally well-calibrated with respect to the surgeon decisions, across groups. This is particularly vital in kidney transplantation, where biased decisions can amplify existing health disparities \citep{rajkomar2018ensuring}. Furthermore, for all surgeon's decisions $z \in \{0, 1\}$, Equation \eqref{Eqn: Sufficiency} can be written as
\begin{equation}
\mathbb{D}_{KL}\Big(P(z \ | \ \hat{y}, x \in \mathcal{X}_m) \ \Big|\Big| \ P(z \ | \ \hat{y}, x \in \mathcal{X}_{m'})\Big) = \displaystyle \sum_{z \in \{0, 1\}} P(z \ | \ \hat{y}, x \in \mathcal{X}_m) \log \frac{P(z \ | \ \hat{y}, x \in \mathcal{X}_m)}{P(z \ | \ \hat{y}, x \in \mathcal{X}_{m'})}.
\label{Eqn: Sufficiency Expansion}
\end{equation}
Here, the term $\pi_{\hat{y}}(z) = P(z \ | \ \hat{y}, x \in \mathcal{X}_m)$ can be expressed using Bayes theorem as follows.
\begin{equation}
\begin{array}{lcl}
P(z \ | \ \hat{y}, x \in \mathcal{X}_m) &=& \displaystyle \frac{P(\hat{y} \ | \ z, x \in \mathcal{X}_m)\cdot P(z \ | \ x \in \mathcal{X}_m)}{P(\hat{y} \ | \ x \in \mathcal{X}_m)}
\\[2ex]
&=& \displaystyle \frac{P(z \ | \ x \in \mathcal{X}_m) \cdot f\Big(\hat{y}_T; k_T(z), \lambda_T(z)\Big) \cdot f\Big(\hat{y}_D; k_D(z), \lambda_D(z)\Big)}{\displaystyle \sum_{\bar{z} \in \{0, 1\}} P(\bar{z} \ | \ x \in \mathcal{X}_m) \cdot f\Big(\hat{y}_T; k_T(\bar{z}), \lambda_T(\bar{z})\Big) \cdot f\Big(\hat{y}_D; k_D(\bar{z}), \lambda_D(\bar{z})\Big)},
\end{array}
\end{equation}
where $f\Big(\hat{y}_T; k_T(z), \lambda_T(z)\Big)$ denote the probability density function (PDF) of the Weibull distribution with shape and scale parameters as $k_T(z), \lambda_T(z)$ for a given $z$.

\begin{figure}[!t]
\centering
\includegraphics[width=\textwidth]{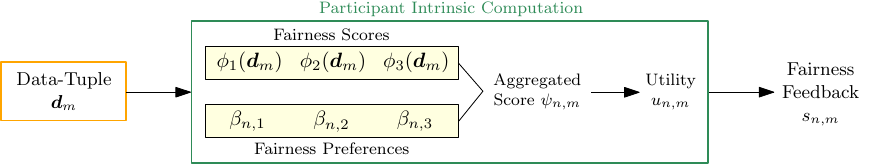}
\caption{Overview of the Proposed Fairness Feedback Model}
\label{Fig: feedback model}
\end{figure}

\subsection{Fairness Feedback Model}
Consider $N$ participants who evaluate the fairness of the UPAT across different social groups. Each participant $n \in \{1, \cdots, N\}$ examines $M$ data-tuples, where an $m^{th}$ data-tuple is defined as $\boldsymbol{d}_m = \{ \boldsymbol{x}_{1:K}^{(m)}, \boldsymbol{z}_{1:K}^{(m)}, \hat{\boldsymbol{y}}_{1:K}^{(m)} \}$, which includes the donor-recipient attributes $\boldsymbol{x}_{1:K}^{(m)}$, surgeon's decisions $\boldsymbol{z}_{1:K}^{(m)}$ and the UPAT's predictions $\hat{\boldsymbol{y}}_{1:K}^{(m)}$ for $K$ donor-recipient pairs. Let $s_{n,m} \in \{1, \cdots, 7\}$ denote the fairness feedback score given by the $n^{th}$ participant for the $m^{th}$ data-tuple, where $s_{n,m} = 1$ denote the system being completely unfair and $s_{n,m} = 7$ as completely fair.

This paper models the participant's fairness feedback based on the assumption that each participant might subconsciously assign weighted preferences over $L$ fairness notions. Let $\boldsymbol{\beta}_n = \{\beta_{n,1}, \cdots, \beta_{n,L}\}$ denote $n^{th}$ participant's  weighted preferences over $L$ notions, where each $\beta_{n,l} \geq 0$ and $\sum_{l=1}^L \beta_{n, l} = 1$ for all $n, l$. Let $\phi_{\ell}(\boldsymbol{d}_m) \in [0, \infty)$ represent the fairness score computed using the $\ell$-th notion for the data-tuple $\boldsymbol{d}_m$. For each data-tuple $\boldsymbol{d}_m$, participant $n$ intrinsically computes $L$ fairness scores corresponding to the $L$ notions. Given that these scores can vary significantly in magnitude and scale, participants normalize these $L$ scores using min-max scaling to facilitate interpretation. Specifically, for each data-tuple $\boldsymbol{d}_m$, the participant performs the following normalization.
\begin{equation} 
\bar{\phi}_l(\boldsymbol{d}_m) = \displaystyle \frac{\phi_l^{\max} - \phi_l(\boldsymbol{d}_m)}{\phi_l^{\max} - \phi_l^{\min}}, 
\label{Eqn: minmax scaling} 
\end{equation}
where $\phi_l^{\min} = \min_l \phi_l(\boldsymbol{d}_m)$ and $\phi_l^{\max} = \max_l \phi_l(\boldsymbol{d}_m)$ are the minimum and maximum fairness scores across all $L$ notions for the data-tuple $d_m$. This transformation maps the scores to a common scale $[0, 1]$. As depicted in Figure \ref{Fig: feedback model}, participant $n$ then aggregates the normalized fairness scores using their preference weights for the data-tuple $d_m$ as
\begin{equation}
\psi_{n,m}(\boldsymbol{\beta}_n) = \sum_{l = 1}^L \beta_{n, l} \cdot \bar{\phi}_l(\boldsymbol{d}_m).
\label{Eqn: participant aggregate fairness evaluation}
\end{equation}
Here, if $\psi_{n,m}(\boldsymbol{\beta}_n)$ is close to 0, participant $n$ perceives the UPAT as a fair system. Conversely, if $\psi_{n,m}(\boldsymbol{\beta}_n)$ is close to 1, UPAT is perceived as unfair. However, it is necessary to encode the aggregated fairness score $\psi_{n,m}(\boldsymbol{\beta}_n) \in [0, 1]$ to 7 point Likert scale. In order to do so, the interval $[0, 1]$ is divided into 7 equal partitions with boundaries defined as $r_i = i/7$ for all $i = 0, 1, \cdots, 7$. Specifically, the region $\mathbb{R}_i$ pertaining to the feedback score $s_{n,m} = i$ is defined as $[r_{7-i}, r_{7-i+1}]$, for all $i = 1, \cdots, 7$. 

Furthermore, assume that each participant's aggregated fairness score $\psi_{n, m}(\boldsymbol{\beta}_n)$ follows Beta distribution $F(\cdot; a_n, b_n)$ with shape parameters defined as $a_n = \psi_{n,m}\cdot c$ and $b_n = (1-\psi_{n,m})\cdot c$, where $c > 0$ is the precision parameter controlling the variance of the distribution. This formulation ensures that the mean of the Beta distribution aligns precisely with the participant's aggregated score, $\mathbb{E}[\psi_{n,m}] = \frac{a_n}{a_n + b_n} = \psi_{n,m}$, effectively centering the distribution at $\psi_{n,m}$. The variance is given by $\text{Var}[\psi_{n,m}] = \frac{\psi_{n,m}(1 - \psi_{n,m})}{c+1}$, which decreases as $c$ increases, reflecting higher confidence and less uncertainty in the participant's assessment. Modeling $\psi_{n,m}$ as a Beta distributed random variable facilitates to capture the inherent uncertainty in participant's perception due to factors such cognitive biases, internal fluctuations, and ambiguity in weighing different fairness notions. This approach aligns with Bayesian principles and psychological theories suggesting that individuals hold probabilistic beliefs rather than fixed values, acknowledging that their judgments are subject to variability and uncertainty. Then, the $n^{th}$ participant experiences a utility $u_{n,m,i}$ as the probability of the intrinsic fairness evaluation $\psi_{n,m}$ to lie in a specific region $\mathbb{R}_i$, for all $i= 1, \cdots, 7$. In other words, the utility that $\psi_{n,m}$ falls in the region $\mathbb{R}_i$ for all $i = 2, \cdots, 7$ is given as
\begin{equation}
\begin{array}{lcl}
u_{n,m,i}(\boldsymbol{d}_m) &=& \displaystyle  F ( r_{7-i+1}; a_n, b_n) - F(r_{7-i}; a_n, b_n)
\\[2ex]
&=& \displaystyle \int_{r_{7-i}}^{r_{7-i+1}} f\Big(r; a_n = \psi_{n,m}\cdot c, b_n =(1-\psi_{n,m})\cdot c \Big) \ dr.
\end{array}
\label{Eqn: utilities}
\end{equation}
In practice, participants often compute a noisy fairness evaluation, due to the ambiguity in their preferences towards different fairness notions. This stochasticity in the preferences across fairness notions can be captured using the mixed-logit model \citep{mcfadden1973conditional, train2009discrete} from discrete choice theory, which is already found to be useful in the field of kidney transplantation \citep{genie2020role, howard2006quality}. Then, based on mixed-logit model, the probability that the $n^{th}$ participant will provide a fairness feedback $i$ is given as
\begin{equation}
P(s_{n,m} = i \ | \ \psi_{n,m} \in \mathbb{R}_i) = \displaystyle \frac{e^{u_{n,m,i}\lambda}}{\sum_{j=1}^7 e^{u_{n,m,j}\lambda}},
\label{Eqn: Mixed logit prob}
\end{equation}
where $\lambda$ is the temperature parameter that captures the participant's sensitivity to the utilities. Hence, the fairness feedback score $s_{n,m}$ is modeled as
\begin{equation}
s_{n,m} = \displaystyle \argmax_i \ P(s_{n,m} = i \ | \ \psi_{n,m} \in \mathbb{R}_i).
\label{Eqn: Estimated Feedback Score}
\end{equation}

\subsection{Fairness Preference Learning Algorithms}
The underlying philosophy of the proposed method is to estimate the optimal social preference weight $\tilde{\boldsymbol{\beta}}^* = \{\tilde{\beta}^*_1, \cdots, \tilde{\beta}^*_L\}$ that minimizes the average feedback regret $\mathcal{L}_F(\tilde{\boldsymbol{\beta}}^*)$, which is given by
\begin{equation}
\mathcal{L}_F(\tilde{\boldsymbol{\beta}}^*) \triangleq \frac{1}{K} \sum_{k=1}^K \left( \frac{1}{N} \sum_{n = 1}^N \left( s_{n, k} - \tilde{s}^*_k \left( \tilde{\boldsymbol{\beta}}^* \right) \right)^2 \right),
\label{Eqn: Feedback Regret}
\end{equation}
where $\tilde{s}^*_m \left( \tilde{\boldsymbol{\beta}}^* \right)$ represents the social fairness evaluation which follows the same definition in Equation \eqref{Eqn: Estimated Feedback Score}, but without having the participant index $n$. 

\begin{algorithm2e}[!t]
\caption{Social Aggregation of Fairness Feedback}
\SetKwFunction{FairnessScores}{FairnessScores}
\SetKwFunction{EstFeedback}{EstFeedback}
\SetKwFunction{VotingRule}{VotingRule}
\label{Alg: SAIPE}
\KwIn{($\boldsymbol{d}_1, s_{1,1}, \cdots, s_{N,1}), \ldots, (\boldsymbol{d}_M, s_{N, 1}, \cdots, s_{N, M})$}
\KwOut{Social preferences $\tilde{\boldsymbol{\beta}}^*$}
Declare learning rate $\delta$\;
Initialize $\tilde{\boldsymbol{\beta}}^{*}$ with $L$-dim. weight
\BlankLine
\For{$e = 1$ \KwTo num\_epochs}
{
\For{$m\leftarrow 1$ \KwTo $M$}
{   
$\phi_1, \ldots, \phi_L \leftarrow \FairnessScores(\boldsymbol{d}_m)$\;
$\tilde{s}_{m} \leftarrow \EstFeedback(\tilde{\boldsymbol{\beta}}^{*(e)}, \phi_1, \ldots, \phi_L)$

$\nabla \ell_m(\tilde{\boldsymbol{\beta}}^{*(e)}) \leftarrow \texttt{SRG}(s_{n, m}, \tilde{s}_{m}, \boldsymbol{\phi}_m, \tilde{\boldsymbol{\beta}}^{*(e)})$\;
}
$\nabla \mathcal{L}_F(\tilde{\boldsymbol{\beta}}^{*(e)}) = \displaystyle \frac{1}{M} \sum_{m = 1}^M \nabla \ell_m(\tilde{\boldsymbol{\beta}}^{*
(e)})$
\\[1ex]
$\tilde{\boldsymbol{\beta}}^{*(e+1)} \leftarrow \mathbb{P} \left[ \tilde{\boldsymbol{\beta}}^{*(e)} - \delta \cdot \nabla \mathcal{L}_F(\tilde{\boldsymbol{\beta}}^{*(e)}) \right]$\
}
\label{Alg: SAFF}
\end{algorithm2e}

The optimal social preference weight $\tilde{\boldsymbol{\beta}}^*$ can be learned using \emph{Social Aggregation of Fairness Feedback} (SAFF) algorithm as shown in Algorithm \ref{Alg: SAFF}, which is developed using projected gradient descent. The projection operator $\mathbb{P}(\cdot)$ ensures that $\tilde{\boldsymbol{\beta}}^*$ is a valid preference weight vector that has entries between 0 to 1 and sums to 1. The regret gradient $\nabla \mathcal{L}_F$ with respect to the unknown preference parameters $\tilde{\boldsymbol{\beta}}^*$ is computed using the well-known \emph{backpropagation} algorithm. Since the feedback regret indirectly depends on the model parameter $\tilde{\boldsymbol{\beta}}^*$, each auxiliary term is expanded until direct dependency is achieved. Hence, the regret gradient $\nabla \mathcal{L}_F$ with respect to the model parameters $\tilde{\boldsymbol{\beta}}^*$ is computed using the following dependency chain of variables:
\begin{equation}
\mathcal{L}_F \stackrel{\text{\eqref{Eqn: Feedback Regret}}}{\longleftarrow} \tilde{s} \stackrel{\text{\eqref{Eqn: Mixed logit prob},\eqref{Eqn: Estimated Feedback Score}}}{\longleftarrow}  \tilde{\boldsymbol{u}} \stackrel{\text{\eqref{Eqn: utilities}}}{\longleftarrow} 
\tilde{\boldsymbol{\psi}} \stackrel{\text{\eqref{Eqn: participant aggregate fairness evaluation}}}{\longleftarrow} \tilde{\boldsymbol{\beta}}^*,
\end{equation}
where the text above the arrows represent the Equation labels corresponding to their respective relationship. Consequently, the gradient of each dependent variable with respect to the model parameter $\tilde{\boldsymbol{\beta}}^*$ has to be computed. Therefore, the backpropagation of gradients is computed using $\texttt{SRG}(\cdot)$, which is defined as
\begin{equation}
\texttt{SRG}\left(s_{n, m}, \tilde{s}_{m}, \boldsymbol{\phi}_m, \tilde{\boldsymbol{\beta}}^{*(e)}\right): \quad \left\{
\begin{array}{rcl}
\displaystyle \nabla_{\boldsymbol{\beta}^*} \mathcal{L}_F &=& \displaystyle ( \nabla_{\tilde{s}} \mathcal{L}_F )^T \cdot \nabla_{\boldsymbol{\tilde{\beta}}^*} \tilde{s}
\\[1ex]
\displaystyle \nabla_{\boldsymbol{\beta}^*} \tilde{s}^* &=& \displaystyle (\nabla_{\tilde{\boldsymbol{u}}} \tilde{s}^* )^T \cdot \nabla_{\tilde{\boldsymbol{\beta}}^*} \tilde{u}
\\[1ex]
\displaystyle \nabla_{\boldsymbol{\beta}^*} \tilde{u} &=& \displaystyle (\nabla_{\tilde{\boldsymbol{\psi}}} \tilde{u})^T \cdot \nabla_{\tilde{\boldsymbol{\beta}}^*} \boldsymbol{\tilde{\psi}},
\end{array} 
\right.
\label{Eqn: regret grad}
\end{equation}
where 
\begin{enumerate}
\item the gradient $\nabla_{\tilde{s}} \mathcal{L}_F$ in the Equation \eqref{Eqn: regret grad} is a $7 \times 1$ vector, that is computed as
\begin{equation}
\nabla_{\tilde{s}} \mathcal{L}_F = -2 \left[ \frac{1}{M} \sum_{m=1}^M \left( \frac{1}{N} \sum_{n = 1}^N \left( s_{n, m} - \tilde{s}^*_m \left( \tilde{\boldsymbol{\beta}}^* \right) \right) \right) \right].
\label{Eqn: regret grad s-tilde* - final}
\end{equation}

\item the gradient $\nabla_{\tilde{\boldsymbol{u}}_{m,i}} \tilde{s}^*$ is a $7 \times 7$ matrix \citep{gao2017properties}, where the $(i,j)^{th}$ entry $\eta_{i,j}$ is given by 
\begin{equation}
\begin{array}{rl}
\eta_{i,j} & = 
\begin{cases}
\displaystyle \frac{\lambda}{\Delta_m^2} \cdot e^{\lambda \tilde{u}_{m,i}} \cdot \sum_{j \neq i} e^{\lambda \tilde{u}_{m,j}}, & \text{if } i = j,
 \\[2ex]
\displaystyle - \frac{\lambda}{\Delta_m^2} \cdot e^{\lambda \tilde{u}_{m,i}} \cdot e^{\lambda \tilde{u}_{m,j}}, & \text{otherwise},
\end{cases}
\end{array}
\label{Eqn: s-tilde* grad u - final}
\end{equation}
with $\Delta_m = \displaystyle \sum_{j = 1}^7 e^{\lambda\cdot \tilde{u}_{m, j}}$ being the normalizing factor and for all $i = j = 1, \cdots, 7$. 

\item the gradient $\nabla_{\boldsymbol{\psi}} u_{m,i}$ in the Equation \eqref{Eqn: regret grad} is a $7 \times 1$ vector, which is computed as
\begin{equation}
\begin{array}{lcl}
\nabla_{\tilde{\boldsymbol{\psi}}} \tilde{u}_{m,i} &=& \displaystyle c \Big[ \left( \displaystyle \frac{\partial}{\partial a}\mathbb{B}(r_{7-i+1}; a, b) - \displaystyle \frac{\partial}{\partial b}\mathbb{B}(r_{7-i+1}; a, b) \right) 
\\[3ex]
&& \qquad - \left( \displaystyle \frac{\partial}{\partial a}\mathbb{B}(r_{7-i}; a, b) - \displaystyle \frac{\partial}{\partial b}\mathbb{B}(r_{7-i}; a, b) \right) \Big],
\label{Eqn: u grad psi - final}
\end{array}
\end{equation}
where $\nabla_{a} \mathbb{B}(r; a, b)$ denotes the gradient of incomplete beta function $\mathbb{B}(\cdot; a, b)$ for some $r$ with respect to the shape parameter $a$. This is solved based on the work of \cite{boik1999derivatives}. The proof for this gradient is detailed in Appendix \ref{Sec: Gradient Proofs}.

\item the gradient $\nabla_{\tilde{\boldsymbol{\beta}}} \tilde{\boldsymbol{\psi}}$ is a $L \times 1$ vector given as
\begin{equation}
\nabla_{\tilde{\boldsymbol{\beta}}} \tilde{\boldsymbol{\psi}} = \{\phi_1(\boldsymbol{d}_m), \cdots, \phi_L(\boldsymbol{d}_m) \}.
\label{Eqn: psi grad beta - final}
\end{equation}
\end{enumerate}
 


\section{Experiment Setup}
The proposed algorithm SAFF is employed on both simulated as well as survey responses. As discussed in Section \ref{Sec: Divergence Measures}, the fairness of the UNOS' Predictive Analytics is measured based on three fairness notions, i.e. $L = 3$. Furthermore, this paper considers the following social groups: \emph{race} = \{Black, All Other Races\}, \emph{gender} = \{Male, Female\}, and \emph{age} = \{$\leq$50, $>$50\}, with advantaged and disadvantaged groups defined as $\mathcal{X}_m$ = \{Other, Male, $\leq$50\} and $\mathcal{X}_{m'}$ = \{Black, Female, $>$50\}, respectively.

For simulation experiments, the social preferences $\tilde{\boldsymbol{\beta}}^*$ are learned across 100 epochs with learning rate declared as $\delta = 0.5$. The complete preprocessed STAR dataset is utilized, which comprises of 624 data-tuples with each tuple containing a single donor profile and 10 recipients matched with this donor. For each experimental setting, a subset of $M \in \{5, 10, 15\}$ data-tuples is randomly sampled from this dataset. This random sampling is performed over 100 iterations, and the outcomes are averaged across these iterations. The experiments are conducted for all participants sizes $N \in \{25, 50, 75, 100\}$, where the true preferences of $N$ participants, $\boldsymbol{\beta}_1, \ldots, \boldsymbol{\beta}_N$, are initialized based on the following scenarios. 

\begin{enumerate}[label=\textbf{Initialization \arabic*:}, leftmargin=20ex, nosep]
\setlength{\itemsep}{1ex}
\vspace{1ex}
\item Random assignment of preferences based on a uniform distribution.
\item Fixed atomic preference, wherein all participants prefer the same fairness notion (e.g., independence).
\item Identical preference, where the participant population is divided such that 33\% prefer independence, 33\% prefer separation, and the remaining 34\% prefer sufficiency.
\end{enumerate}


In the survey experiment, the initial estimate of the social preference, $\tilde{\boldsymbol{\beta}}^{*(0)}$, is randomly generated from a uniform distribution. Participants rate the fairness of the Predictive Analytics using a 7-point Likert scale, i.e., $\boldsymbol{s}_n \in \{1, 2, \ldots, 7\}$. The social preferences are learned over 100 epochs using a learning rate of $\delta = 0.5$ with $M = 10$ data-tuples, each tuple containing $K = 10$ donor-recipient pairs presented to $N = 75$ participants.

\section{Results}

\begin{figure}[!t]
\centering
\includegraphics[width=\textwidth]{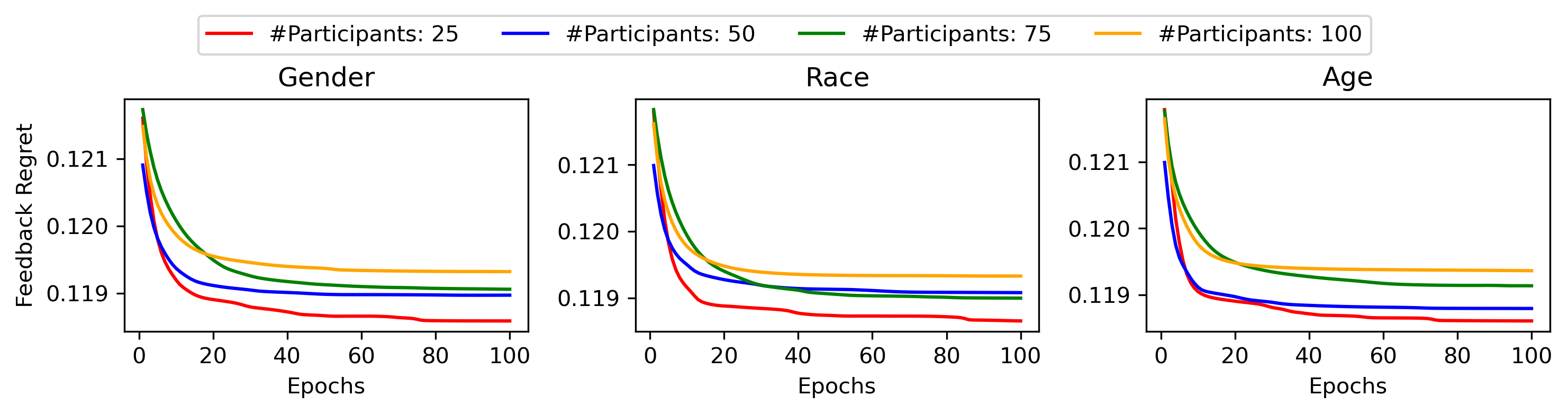}
\caption{Convergence of Feedback Regret in Simulation with $M = 10$ Tuples.}
\label{Fig: Simulation results case1; 10 tuples}
\vspace{-2ex}
\end{figure}


\subsection{Simulation Results}
Figure \ref{Fig: Simulation results case1; 10 tuples} illustrates the social feedback regret for varying participant counts, $N = \ $\{25, 50, 75, 100\}, where each participant receives $M = 10$ data tuples. The preferences are generated following the Initialization 1 procedure. The left-most plot shows the convergence of social feedback regret across gender, derived from responses to Question Q4. Similarly, the middle plot presents social feedback regret as a function of race, based on responses to Question Q3, while the right-most plot illustrates the regret associated with age, as determined by responses to Question Q2 (refer to Figure \ref{Fig: Survey Questions}). The results indicate that preference regret consistently decreases as the number of epochs increases, regardless of the sensitive attribute considered or the specific combination of data tuple size and participant count. However, increasing the number of participants and/or the data tuple size has only a marginal impact on further reducing social feedback regret. Additionally, similar regret convergence rates are observed when evaluated for all tuple sizes $M = \{5, 10, 15\}$ across all social groups, as depicted in Figure \ref{Fig: Feedback regret - case1} (Appendix \ref{App: More results}). These results validate our hypothesis that participant preferences can be effectively modeled using a Mixed-Logit framework. The relatively low feedback regret observed (Figure \ref{Fig: Feedback regret - case1}) suggests that the proposed model accurately estimates social preferences that align with individual participant preferences.  

To further assess the proposed learning algorithm, participant preferences were initialized using Initialization scenarios 2 and 3. In the first scenario, participants' true preferences were configured to exclusively prefer the independence fairness notion, i.e., $\boldsymbol{\beta}_1 = \cdots = \boldsymbol{\beta}_N = \{1, 0, 0\}$. Under the same experimental setup as before, the learned social preferences exhibited a strong preference (approximately 0.95) toward independence, with negligible weights assigned to sufficiency and separation notions. These findings confirm that the proposed learning algorithm functions as intended, effectively capturing and reflecting the specified preference structures. With Initialization 3, the results show inconsistency in socially preferred notion across algorithm runs. This behavior arises due to the absence of a clear majority, creating a loss landscape with multiple minima where no single social ranking distinctly minimizes the loss function.

\begin{table}[!t]
\caption{Survey Experiment Outcomes}
\centering
\begin{tabular}{l c c c c c c c c c}
\toprule
& \multicolumn{3}{c}{\textbf{Fairness Scores}} & \multicolumn{5}{c}{\textbf{Social Fairness}} & \textbf{Social Fairness}\\
&\multicolumn{3}{c}{} & \multicolumn{5}{c}{\textbf{Preferences}} & \textbf{Feedback Score}
\\[1ex]
& $\phi_1$ & $\phi_2$ & $\phi_3$ && $\tilde{\beta}^*_1$ & $\tilde{\beta}^*_2$ & $\tilde{\beta}^*_3$ & & $\tilde{s}^*$\\
\midrule
Gender & 0.09 & 0.30 & 0.01 && 0.29 & 0.31 & \textbf{0.40} & & 7 (completely fair) \\
Race & 0.13 & 0.23 & 0.02 && 0.14 & \textbf{0.44} & 0.42 & & 7 (completely fair)\\
Age & 4.10 & 3.66 & 1.08 && 0.27 & \textbf{0.39} & 0.34 & & 1 (completely unfair)\\
\bottomrule
\end{tabular}
\label{Tab: survey outcomes}
\vspace{-2ex}
\end{table}

\subsection{Survey Results}
Table \ref{Tab: survey outcomes} demonstrates the initial fairness assessment of UPAT, which is computed based on 10 data-tuples presented in the survey experiment. The attribute Age exhibits substantially elevated disparities across all fairness dimensions (Independence: 4.10, Separation: 3.66, Sufficiency: 1.08). The attributes Gender and Race demonstrate relatively minimal independence (0.09, 0.13) and sufficiency (0.01, 0.02) disparities, suggesting the model's predictions maintain similar distributions and outcome independence across these demographics. However, the elevated separation metrics for both Gender (0.30) and Race (0.23) indicate differential error rates conditional on the surgeon's decisions. Moreover, as shown in Table \ref{Tab: survey outcomes}, the recruited participants exhibit a strong preference for fairness notions that are dependent on surgeon decisions, with higher preference weights for separation ($\tilde{\beta}_2^* = 0.31$ for gender, $0.44$ for race, $0.39$ for age) and sufficiency ($\tilde{\beta}_3^* = 0.40$ for gender, $0.42$ for race, $0.34$ for age) compared to independence ($\tilde{\beta}_1^* = 0.29$ for gender, $0.14$ for race, $0.27$ for age).  


\section{Discussion on Survey Results} 
Table \ref{Tab: survey outcomes} presents significantly high fairness scores (a.k.a. $\phi_1 = 4.10$, $\phi_2 = 3.66$ and $\phi_3 = 1.08$) across young and old recipients. Consequently, the proposed algorithm \emph{SAFF} naturally estimated the social fairness feedback score from public feedback as $\tilde{s}^* = 1$ and is therefore deemed unfair based on age. However, this disparity of kidney offers between young and old recipients is justified in clinical practice, as age-stratified risk prediction is fundamental to organ allocation protocols \citep{rana2015survival}. Age is one of the strongest independent predictors of post-transplant outcomes and waitlist mortality, given its role in allocation scores such as Kidney Donor Profile Index (KDPI) and Estimated Post-Transplant Survival (EPTS) \citep{massie2016risk, clayton2014external}. In addition, age-based risk stratification is also justified as long as the organ utilization is maximized \citep{reese2015living, ross2012equal}. Despite these justifications, the public stakeholders find UPAT to be unfair in terms of age, which leads to distrust. Therefore, UNOS is strongly advised to improve public awareness regarding age-stratified risks in kidney placement and why such a discrimination exhibited by UPAT is justified clinically.

Furthermore, large preference weights for $\tilde{\beta}_2^*$ and $\tilde{\beta}_3^*$ in the case of race indicate that the public stakeholders exhibit higher preference towards sufficiency and separation based fairness scores. Therefore, it is imperative that UNOS ensures that $\phi_2$ and $\phi_3$ stay low to maintain public trust. Although $\phi_3$ is relatively close to zero, $\phi_2$ is not necessarily insignificant and can potentially lead to social distrust, if not monitored in the future. Additionally, since $\tilde{\beta}_1^* = 0.14$ (low value) for race, public participants tend to condition their fairness evaluations on surgeon decisions with high probability. This demonstrates that the participants clearly understood the kidney transplant pipeline wherein surgeons make a decision after seeing UPAT's analysis regarding any given offer.

On the contrary, a uniform preference is exhibited by public stakeholders across independence, sufficiency and separation based fairness scores in the case of gender. Given the low fairness scores, the social fairness feedback score is estimated as $\tilde{s}^* = 7$, which means the public stakeholders perceive that UPAT is fair with respect to both gender and race. However, it is important to note that the separation score $\phi_2$ is not insignificant in the case of gender. If this value raises in the future, it can deteriorate public trust in UPAT. 


\paragraph{Limitations} The limitation of the proposed approach lies in the assumption that the fairness notions are computed using Weibull distribution, since UPAT is based on Cox-Hazards model. Although Weibull distribution has been used extensively in the kidney transplantation literature \citep{hashemian2013comparison, yahav2010predicting}, there could be some practical conditions under which Weibull distribution does not fit the model predictions. Furthermore, it is frequently observed in several other healthcare applications such as breast cancer survival analysis \citep{baghestani2016survival}, COVID-19 data modeling \citep{imran2024development}, aging and disease patterns \citep{matsushita1992lifetime}, and modeling censored medical datasets \citep{feroze2022applicability}. Additionally, the demographics of our study participants differ from U.S. census statistics, which may impact the generalizability of our findings. In particular, Hispanic participants are underrepresented in our survey (3.4\%) compared to their census proportion (18\%). Conversely, our survey sample includes a higher proportion of Black (19\% vs. 12\%) and Asian (12\% vs. 5.6\%) participants than the census. Furthermore, the educational background of our respondents also differs significantly from census data, with a much higher proportion of individuals holding a Bachelor's degree (40\% vs. 20\%) and fewer participants with only a high school education (18\% vs. 26.5\%). 

\paragraph{Planned Future Work} In future work, we aim to develop novel survey experiments tailored to collect feedback from individual stakeholder groups based on their expertise within different stages of the kidney placement pipeline. For example, transplant surgeons can assess UPAT's fairness by evaluating medical attributes of donors and recipients, such as diabetes status and glomerular filtration rate, which differs from the public survey presented in this study. Additionally, transplant surgeons can provide an independent assessment of mortality rate predictions for a given donor-recipient pair as part of the survey feedback.


\bibliography{sample-bibliography}

\begin{thebibliography}{43}
\providecommand{\natexlab}[1]{#1}
\providecommand{\url}[1]{\texttt{#1}}
\expandafter\ifx\csname urlstyle\endcsname\relax
  \providecommand{\doi}[1]{doi: #1}\else
  \providecommand{\doi}{doi: \begingroup \urlstyle{rm}\Url}\fi

\bibitem[Agarwal et~al.(2019)Agarwal, Dud{\'\i}k, and Wu]{agarwal2019fair}
Alekh Agarwal, Miroslav Dud{\'\i}k, and Zhiwei~Steven Wu.
\newblock {Fair Regression: Quantitative Definitions and Reduction-based Algorithms}.
\newblock In \emph{International Conference on Machine Learning}, pages 120--129. PMLR, 2019.

\bibitem[Baghestani et~al.(2016)Baghestani, Moghaddam, Majd, Akbari, Nafissi, and Gohari]{baghestani2016survival}
Ahmad~Reza Baghestani, Sahar~Saeedi Moghaddam, Hamid~Alavi Majd, Mohammad~Esmaeil Akbari, Nahid Nafissi, and Kimiya Gohari.
\newblock {Survival Analysis of Patients with Breast Cancer using Weibull Parametric Model}.
\newblock \emph{Asian Pacific Journal of Cancer Prevention}, 16\penalty0 (18):\penalty0 8567--8571, 2016.

\bibitem[Barocas et~al.(2019)Barocas, Hardt, and Narayanan]{barocas2019fairness}
Solon Barocas, Moritz Hardt, and Arvind Narayanan.
\newblock Fairness and machine learning. fairmlbook. org, 2019.

\bibitem[Bauckhage and Manshaei(2014)]{bauckhage2014kernel}
Christian Bauckhage and Kasra Manshaei.
\newblock Kernel archetypal analysis for clustering web search frequency time series.
\newblock In \emph{2014 22nd International Conference on Pattern Recognition}, pages 1544--1549. IEEE, 2014.

\bibitem[Bekbolsynov et~al.(2022)Bekbolsynov, Mierzejewska, Khuder, Ekwenna, Rees, Green, and Stepkowski]{bekbolsynov2022improving}
Dulat Bekbolsynov, Beata Mierzejewska, Sadik Khuder, Obinna Ekwenna, Michael Rees, Robert~C Green, and Stanislaw~M Stepkowski.
\newblock {Improving Access to HLA-matched Kidney Transplants for African American Patients}.
\newblock \emph{Frontiers in Immunology}, 13:\penalty0 832488, 2022.

\bibitem[Boik and Robinson-Cox(1999)]{boik1999derivatives}
Robert~J Boik and James~F Robinson-Cox.
\newblock Derivatives of the incomplete beta function.
\newblock \emph{Journal of Statistical Software}, 3:\penalty0 1--20, 1999.

\bibitem[Bureau(2021)]{Census2021ACS}
U.S.~Census Bureau.
\newblock {ACS Demographic and Housing Estimates}.
\newblock U.S. Census Bureau, 2021.

\bibitem[Charalambides(2021)]{charalambidesProlific}
Nick Charalambides.
\newblock {We Recently Went Viral on TikTok - Here’s What We Learned}.
\newblock August 2021.

\bibitem[Chouldechova(2017)]{chouldechova2017fair}
Alexandra Chouldechova.
\newblock {Fair Prediction with Disparate Impact: A Study of Bias in Recidivism Prediction Instruments}.
\newblock \emph{Big data}, 5\penalty0 (2):\penalty0 153--163, 2017.

\bibitem[Chouldechova and Roth(2018)]{chouldechova2018frontiers}
Alexandra Chouldechova and Aaron Roth.
\newblock {The Frontiers of Fairness in Machine Learning}.
\newblock \emph{arXiv preprint arXiv:1810.08810}, 2018.

\bibitem[Chzhen et~al.(2020)Chzhen, Denis, Hebiri, Oneto, and Pontil]{chzhen2020fair}
Evgenii Chzhen, Christophe Denis, Mohamed Hebiri, Luca Oneto, and Massimiliano Pontil.
\newblock {Fair Regression with Wasserstein Barycenters}.
\newblock \emph{Advances in Neural Information Processing Systems}, 33:\penalty0 7321--7331, 2020.

\bibitem[Clayton et~al.(2014)Clayton, McDonald, Snyder, Salkowski, and Chadban]{clayton2014external}
PA~Clayton, SP~McDonald, JJ~Snyder, N~Salkowski, and SJ~Chadban.
\newblock {External Validation of the Estimated Posttransplant Survival Score for Allocation of Deceased Donor Kidneys in the United States}.
\newblock \emph{American Journal of Transplantation}, 14\penalty0 (8):\penalty0 1922--1926, 2014.

\bibitem[Feroze et~al.(2022)Feroze, Tahir, Noor-ul Amin, Nisar, Alqahtani, Abbas, Ali, and Jirawattanapanit]{feroze2022applicability}
Navid Feroze, Uroosa Tahir, Muhammad Noor-ul Amin, Kottakkaran~Sooppy Nisar, Mohammed~S Alqahtani, Mohamed Abbas, Rashid Ali, and Anuwat Jirawattanapanit.
\newblock {Applicability of Modified Weibull Extension Distribution in Modeling Censored Medical Datasets: A Bayesian Perspective}.
\newblock \emph{Scientific Reports}, 12\penalty0 (1):\penalty0 17157, 2022.

\bibitem[Gao and Pavel(2017)]{gao2017properties}
Bolin Gao and Lacra Pavel.
\newblock {On the Properties of the Softmax Function with Application in Game Theory and Reinforcement Learning}.
\newblock \emph{arXiv preprint arXiv:1704.00805}, 2017.

\bibitem[Genie et~al.(2020)Genie, Nicol{\'o}, and Pasini]{genie2020role}
Mesfin~G Genie, Antonio Nicol{\'o}, and Giacomo Pasini.
\newblock {The Role of Heterogeneity of Patients’ Preferences in Kidney Transplantation}.
\newblock \emph{Journal of Health Economics}, 72:\penalty0 102331, 2020.

\bibitem[Gotlieb et~al.(2022)Gotlieb, Azhie, Sharma, Spann, Suo, Tran, Orchanian-Cheff, Wang, Goldenberg, Chass{\'e}, et~al.]{gotlieb2022promise}
Neta Gotlieb, Amirhossein Azhie, Divya Sharma, Ashley Spann, Nan-Ji Suo, Jason Tran, Ani Orchanian-Cheff, Bo~Wang, Anna Goldenberg, Michael Chass{\'e}, et~al.
\newblock The promise of machine learning applications in solid organ transplantation.
\newblock \emph{{NPJ Digital Medicine}}, 5\penalty0 (1):\penalty0 89, 2022.

\bibitem[Harding et~al.(2023)Harding, Di, Pastan, Rossi, DuBay, Gompers, and Patzer]{harding2023sex}
Jessica~L Harding, Mengyu Di, Stephen~O Pastan, Ana Rossi, Derek DuBay, Annika Gompers, and Rachel~E Patzer.
\newblock {Sex/Gender-based Disparities in Early Transplant Access by Attributed Cause of Kidney Disease - Evidence from a Multiregional Cohort in the Southeast United States}.
\newblock \emph{Kidney International Reports}, 8\penalty0 (12):\penalty0 2580--2591, 2023.

\bibitem[Hardt et~al.(2016)Hardt, Price, and Srebro]{MoritzOpportunities}
M.~Hardt, E.~Price, and N.~Srebro.
\newblock {Equality of Opportunity in Supervised Learning}.
\newblock In D.~D. Lee, M.~Sugiyama, U.~V. Luxburg, I.~Guyon, and R.~Garnett, editors, \emph{Advances in Neural Information Processing Systems 29}, pages 3315--3323. Curran Associates, Inc., 2016.

\bibitem[Harrison et~al.(2020)Harrison, Hanson, Jacinto, Ramirez, and Ur]{harrison2020empirical}
Galen Harrison, Julia Hanson, Christine Jacinto, Julio Ramirez, and Blase Ur.
\newblock {An Empirical Study on the Perceived Fairness of Realistic, Imperfect Machine Learning Models}.
\newblock In \emph{Proceedings of the 2020 Conference on Fairness, Accountability, and Transparency}, pages 392--402, 2020.

\bibitem[Hashemian et~al.(2013)Hashemian, Beiranvand, Rezaei, Reissi, et~al.]{hashemian2013comparison}
Amir~Hossein Hashemian, Behrouz Beiranvand, Mansour Rezaei, Dariush Reissi, et~al.
\newblock {A Comparison between Cox Regression and Parametric Methods in Analyzing Kidney Transplant Survival}.
\newblock \emph{World Applied Sciences Journal}, 26\penalty0 (4):\penalty0 502--7, 2013.

\bibitem[Howard(2006)]{howard2006quality}
David~H Howard.
\newblock {Quality and Consumer Choice in Healthcare: Evidence from Kidney Transplantation}.
\newblock \emph{The BE Journal of Economic Analysis \& Policy}, 5\penalty0 (1):\penalty0 0000101515153806531349, 2006.

\bibitem[Imran et~al.(2024)Imran, Alsadat, Tahir, Jamal, Elgarhy, Ahmad, and Johannssen]{imran2024development}
Muhammad Imran, Najwan Alsadat, MH~Tahir, Farrukh Jamal, Mohammed Elgarhy, Hijaz Ahmad, and Arne Johannssen.
\newblock {The Development of an Extended Weibull Model with Applications to Medicine, Industry and Actuarial Sciences}.
\newblock \emph{Scientific Reports}, 14\penalty0 (1):\penalty0 12338, 2024.

\bibitem[Kleinberg et~al.(2017)Kleinberg, Mullainathan, and Raghavan]{kleinberg2016inherent}
Jon Kleinberg, Sendhil Mullainathan, and Manish Raghavan.
\newblock {Inherent Trade-Offs in the Fair Determination of Risk Scores}.
\newblock \emph{Innovations in Theoretical Computer Science (ITCS) Conference}, 2017.

\bibitem[Lavanchy et~al.(2023)Lavanchy, Reichert, Narayanan, and Savani]{lavanchy2023applicants}
Maude Lavanchy, Patrick Reichert, Jayanth Narayanan, and Krishna Savani.
\newblock {Applicants’ Fairness Perceptions of Algorithm-Driven Hiring Procedures}.
\newblock \emph{Journal of Business Ethics}, pages 1--26, 2023.

\bibitem[Magee et~al.(2004)Magee, Bucuvalas, Farmer, Harmon, Hulbert-Shearon, and Mendeloff]{magee2004pediatric}
John~C Magee, John~C Bucuvalas, Douglas~G Farmer, William~E Harmon, Tempie~E Hulbert-Shearon, and Eric~N Mendeloff.
\newblock {Pediatric Transplantation}.
\newblock \emph{American Journal of Transplantation}, 4:\penalty0 54--71, 2004.

\bibitem[Massie et~al.(2016)Massie, Leanza, Fahmy, Chow, Desai, Luo, King, Bowring, and Segev]{massie2016risk}
Allan~B Massie, Joseph Leanza, Lara~M Fahmy, Eric~KH Chow, Niraj~M Desai, Xun Luo, Elizabeth~A King, Mary~G Bowring, and Dorry~L Segev.
\newblock {A Risk Index for Living Donor Kidney Transplantation}.
\newblock \emph{American Journal of Transplantation}, 16\penalty0 (7):\penalty0 2077--2084, 2016.

\bibitem[Matsushita et~al.(1992)Matsushita, Hagiwara, Shiota, Shimada, Kuramoto, and Toyokura]{matsushita1992lifetime}
Satoru Matsushita, Kouichi Hagiwara, Takahiro Shiota, Hiroyuki Shimada, Kizuku Kuramoto, and Yasuo Toyokura.
\newblock {Lifetime Data Analysis of Disease and Aging by the Weibull Probability Distribution}.
\newblock \emph{{Journal of Clinical Epidemiology}}, 45\penalty0 (10):\penalty0 1165--1175, 1992.

\bibitem[McCulloh et~al.(2023)McCulloh, Stewart, Kiernan, Yazicioglu, Patsolic, Zinner, Mohan, and Cartwright]{mcculloh2023experiment}
Ian McCulloh, Darren Stewart, Kevin Kiernan, Ferben Yazicioglu, Heather Patsolic, Christopher Zinner, Sumit Mohan, and Laura Cartwright.
\newblock {An Experiment on the Impact of Predictive Analytics on Kidney Offer Acceptance Decisions}.
\newblock \emph{American Journal of Transplantation}, 23:\penalty0 957--965, 2023.

\bibitem[McFadden et~al.(1973)]{mcfadden1973conditional}
Daniel McFadden et~al.
\newblock {Conditional Logit Analysis of Qualitative Choice Behavior}.
\newblock \emph{Frontiers in Econometrics}, pages 105--142, 1973.

\bibitem[Rajkomar et~al.(2018)Rajkomar, Hardt, Howell, Corrado, and Chin]{rajkomar2018ensuring}
Alvin Rajkomar, Michaela Hardt, Michael~D Howell, Greg Corrado, and Marshall~H Chin.
\newblock {Ensuring Fairness in Machine Learning to Advance Health Equity}.
\newblock \emph{{Annals of Internal Medicine}}, 169\penalty0 (12):\penalty0 866--872, 2018.

\bibitem[Rana et~al.(2015)Rana, Gruessner, Agopian, Khalpey, Riaz, Kaplan, Halazun, Busuttil, and Gruessner]{rana2015survival}
Abbas Rana, Angelika Gruessner, Vatche~G Agopian, Zain Khalpey, Irbaz~B Riaz, Bruce Kaplan, Karim~J Halazun, Ronald~W Busuttil, and Rainer~WG Gruessner.
\newblock {Survival Benefit of Solid-Organ Transplant in the United States}.
\newblock \emph{JAMA Surgery}, 150\penalty0 (3):\penalty0 252--259, 2015.

\bibitem[Reese et~al.(2015)Reese, Boudville, and Garg]{reese2015living}
Peter~P Reese, Neil Boudville, and Amit~X Garg.
\newblock {Living Kidney Donation: Outcomes, Ethics, and Uncertainty}.
\newblock \emph{The Lancet}, 385\penalty0 (9981):\penalty0 2003--2013, 2015.

\bibitem[Ross et~al.(2012)Ross, Parker, Veatch, Gentry, and Thistlethwaite~Jr]{ross2012equal}
Lainie~F Ross, William Parker, Robert~M Veatch, Sommer~E Gentry, and JR~Thistlethwaite~Jr.
\newblock {Equal Opportunity Supplemented by Fair Innings: Equity and Efficiency in Allocating Deceased Donor Kidneys}.
\newblock \emph{American Journal of Transplantation}, 12\penalty0 (8):\penalty0 2115--2124, 2012.

\bibitem[Salas et~al.(2022)Salas, Chua, Rossi, Shah, Katz-Greenberg, Coscia, Sawinski, and Adey]{salas2022sex}
Maria Aurora~Posadas Salas, Elizabeth Chua, Ana Rossi, Silvi Shah, Goni Katz-Greenberg, Lisa Coscia, Deirdre Sawinski, and Deborah Adey.
\newblock {Sex and Gender Disparity in Kidney Transplantation: Historical and Future Perspectives}.
\newblock \emph{{Clinical Transplantation}}, 36\penalty0 (12):\penalty0 e14814, 2022.

\bibitem[Shipe et~al.(2019)Shipe, Deppen, Farjah, and Grogan]{shipe2019developing}
Maren~E Shipe, Stephen~A Deppen, Farhood Farjah, and Eric~L Grogan.
\newblock Developing prediction models for clinical use using logistic regression: An overview.
\newblock \emph{Journal of thoracic disease}, 11\penalty0 (Suppl 4):\penalty0 S574, 2019.

\bibitem[Srivastava et~al.(2019)Srivastava, Heidari, and Krause]{srivastava2019mathematical}
Megha Srivastava, Hoda Heidari, and Andreas Krause.
\newblock {Mathematical Notions vs. Human Perception of Fairness: A Descriptive Approach to Fairness for Machine Learning}.
\newblock In \emph{Proceedings of the 25th ACM SIGKDD International Conference on Knowledge Discovery \& Data Mining}, pages 2459--2468, 2019.

\bibitem[Steinberg et~al.(2020)Steinberg, Reid, and O'Callaghan]{steinberg2020fairness}
Daniel Steinberg, Alistair Reid, and Simon O'Callaghan.
\newblock Fairness measures for regression via probabilistic classification.
\newblock \emph{arXiv preprint arXiv:2001.06089}, 2020.

\bibitem[Telukunta et~al.(2024)Telukunta, Rao, Stickney, Nadendla, and Canfield]{telukunta2024fairpreference}
Mukund Telukunta, Sukruth Rao, Gabriella Stickney, Venkata Sriram~Siddhardh Nadendla, and Casey Canfield.
\newblock {Learning Social Fairness Preferences from Non-Expert Stakeholder Opinions in Kidney Placement}.
\newblock In Tom Pollard, Edward Choi, Pankhuri Singhal, Michael Hughes, Elena Sizikova, Bobak Mortazavi, Irene Chen, Fei Wang, Tasmie Sarker, Matthew McDermott, and Marzyeh Ghassemi, editors, \emph{Proceedings of the fifth Conference on Health, Inference, and Learning}, volume 248 of \emph{Proceedings of Machine Learning Research}, pages 683--695. PMLR, 27--28 Jun 2024.

\bibitem[Train(2009)]{train2009discrete}
Kenneth~E Train.
\newblock \emph{Discrete choice methods with simulation}.
\newblock Cambridge university press, 2009.

\bibitem[Turner et~al.(2020)Turner, Engelsma, Taylor, Sharma, and Demiris]{turner2020recruiting}
Anne~M Turner, Thomas Engelsma, Jean~O Taylor, Rashmi~K Sharma, and George Demiris.
\newblock {Recruiting Older Adult Participants through Crowdsourcing Platforms: Mechanical Turk versus Prolific Academic}.
\newblock In \emph{AMIA Annual Symposium Proceedings}, volume 2020, page 1230. American Medical Informatics Association, 2020.

\bibitem[Veinot et~al.(2018)Veinot, Mitchell, and Ancker]{veinot2018good}
Tiffany~C Veinot, Hannah Mitchell, and Jessica~S Ancker.
\newblock {Good Intentions are Not Enough: How Informatics Interventions Can Worsen Inequality}.
\newblock \emph{Journal of the American Medical Informatics Association}, 25\penalty0 (8):\penalty0 1080--1088, 2018.

\bibitem[Veroux et~al.(2012)Veroux, Grosso, Corona, Mistretta, Giaquinta, Giuffrida, Sinagra, and Veroux]{veroux2012age}
Massimiliano Veroux, Giuseppe Grosso, Daniela Corona, Antonio Mistretta, Alessia Giaquinta, Giuseppe Giuffrida, Nunzia Sinagra, and Pierfrancesco Veroux.
\newblock {Age is an Important Predictor of Kidney Transplantation Outcome}.
\newblock \emph{Nephrology Dialysis Transplantation}, 27\penalty0 (4):\penalty0 1663--1671, 2012.

\bibitem[Yahav and Shmueli(2010)]{yahav2010predicting}
Inbal Yahav and Galit Shmueli.
\newblock {Predicting Potential Survival Rates of Kidney Transplant Candidates from Databases with Existing Allocation Policies}.
\newblock In \emph{Proceedings of the 5th INFORMS Workshop on Data Mining and Health Informatics (DM-HI 2010), Austin, TX}, 2010.

\end{thebibliography}

\newpage
\appendix

\section{Survey Information \label{App: Complete Survey Design}}
\begin{figure}[!h]
\centering
\includegraphics[width=\textwidth]{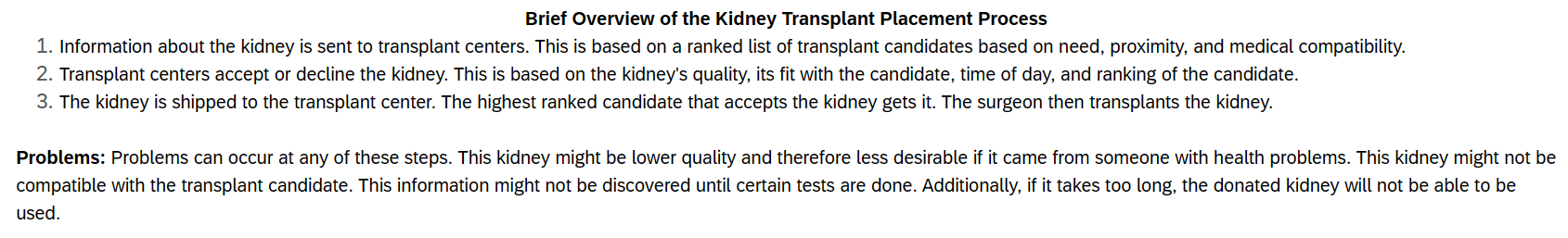}
\includegraphics[width=\textwidth]{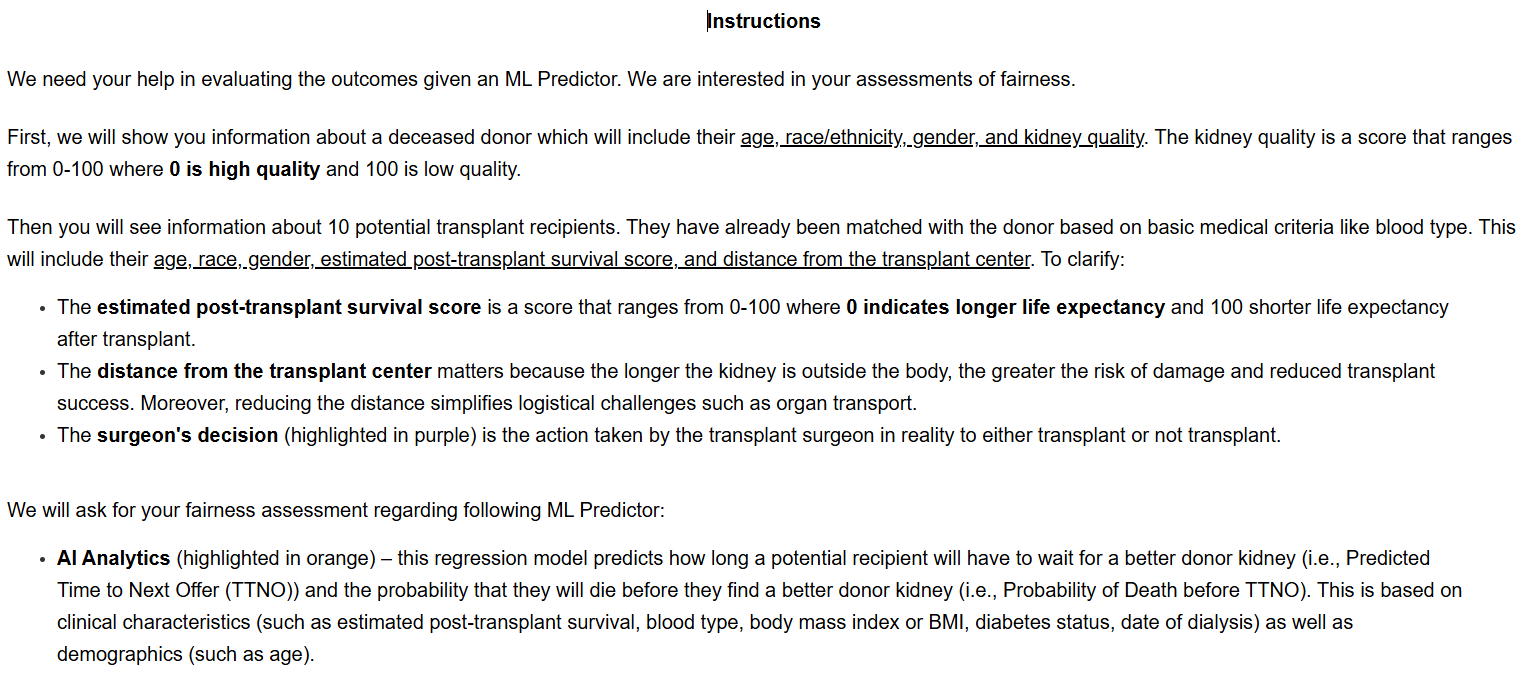}
\caption{Survey Instructions Presented to the Participants.}
\label{Fig: Survey Instructions}
\end{figure}

As shown in Figure \ref{Fig: Survey Instructions}, the recruited participants are first presented with a brief overview of the kidney placement process in the United States which includes information regarding the transplant centers, kidney offers, identifying potential recipient, and transportation of the donor kidney. Following this, an attention check question is presented to the participants: Which of the following is not a criteria to rank list of transplant candidates?. In the next page, instructions regarding the survey experiment is detailed. Specifically, this page explains how the data-tuple is represented, different donor-recipient attributes involving in a data-tuple, and what is expected from the participants.

\section{Gradient Computation Proofs \label{Sec: Gradient Proofs}}

\subsection{Computation of $\nabla_{\psi}u$}
From the definition of participant utilities in Equation \eqref{Eqn: utilities},
\begin{equation}
\frac{\partial u_{n,m,i}}{\partial \psi_{n,m}} = \frac{\partial}{\partial \psi_{n,m}} \Big[  F ( r_{7-i+1}; a_n = \psi_{n,m}\cdot c, b_n = (1-\psi_{n,m})\cdot c) - F(r_{7-i}; a_n, b_n) \Big]
\end{equation}
Generalize $F( r_{7-i+1}; a_n = \psi_{n,m}\cdot c, b_n = (1-\psi_{n,m})\cdot c)$ for some $r$. The CDF of the Beta distribution can be written as
\begin{equation}
\begin{array}{lcl}
\displaystyle \frac{\partial}{\partial \psi_{n,m}} F(r; a_n, b_n) &=& \displaystyle \frac{\partial}{\partial \psi_{n,m}} \left[ \frac{\mathbb{B}(r; a_n, b_n)}{\mathbb{B}(a_n, b_n)} \right] 
\\[3ex]
&=& \displaystyle \frac{\mathbb{B}(a_n, b_n) \cdot \frac{\partial}{\partial \psi_{n,m}} \left[ \mathbb{B}(r; a_n, b_n) \right] -  \mathbb{B}(r; a_n, b_n) \cdot \frac{\partial}{\partial \psi_{n,m}} \left[\mathbb{B}(a_n, b_n) \right] }{\mathbb{B}(a_n, b_n)^2}
\\[3ex]
&=& \displaystyle \frac{1}{\mathbb{B}(a_n, b_n)} \cdot \frac{\partial}{\partial \psi_{n,m}} \left[ \mathbb{B}(r; a_n, b_n) \right] -  \frac{\mathbb{I}(r;a_n, b_n)}{\mathbb{B}(a_n, b_n)} \cdot \frac{\partial}{\partial \psi_{n,m}} \left[\mathbb{B}(a_n, b_n) \right],
\label{Eqn: Beta CDF Expansion}
\end{array}
\end{equation}
where $\mathbb{I}(r;a_n, b_n) = \frac{\mathbb{B}(r; a_n, b_n)}{\mathbb{B}(a_n, b_n)}$ is the regularized incomplete beta function and $\mathbb{B}(r; a_n, b_n)$ incomplete beta function. First, differentiate the term $\mathbb{B}(r; a_n, b_n)$ with respect to $\psi_{n,m}$ using the definition of CDF of Beta distribution.
\begin{equation}
\begin{array}{lcl}
\displaystyle \frac{\partial}{\partial \psi_{n,m}} \left[ \mathbb{B}(r; a_n, b_n) \right] &=& \displaystyle \int_{0}^r \frac{\partial}{\partial \psi_{n,m}} \left[  t^{a_n-1} (1 - t)^{b_n-1}\cdot dt \right]
\\[3ex]
&=& \displaystyle \int_{0}^r \left[ t^{a_n-1} \cdot \underbrace{\frac{\partial}{\partial \psi_{n,m}}(1 - t)^{b_n-1}}_{B1} + (1 - t)^{b_n-1}\cdot\underbrace{\frac{\partial}{\partial \psi_{n,m}}t^{a_n-1}}_{B2} \right]\cdot dt
\label{Eqn: Incomplete Beta wrt psi}
\end{array}
\end{equation}
Solving the term $B1$: Let $z = (1 - t)^{b_n-1} = t^{(1-\psi_{n,m})c-1}$. Apply $\log$ and differentiate both sides with respect to $\psi_{n,m}$.
\begin{equation}
\begin{array}{rcl}
\displaystyle \frac{1}{z} \frac{\partial z}{\partial \psi_{n,m}} &=& \log(1-t)\cdot (-c)
\\[1ex]
\displaystyle \frac{\partial}{\partial \psi_{n,m}} \left[ (1 - t)^{b_n-1} \right] &=& -\log(1-t)\cdot c\cdot (1 - t)^{b_n-1}
\label{Eqn: Incomplete beta - B1}
\end{array}
\end{equation}
Solving the term $B2$: Let $z = t^{a_n-1} = t^{\psi_{n,m}c-1}$. Apply $\log$ and differentiate both sides with respect to $\psi_{n,m}$.
\begin{equation}
\begin{array}{rcl}
\displaystyle \frac{1}{z} \frac{\partial z}{\partial \psi_{n,m}} &=& \log t\cdot (c)
\\[1ex]
\displaystyle \frac{\partial}{\partial \psi_{n,m}} \left[ (1 - t)^{b_n-1} \right] &=&\log t\cdot c\cdot t^{a_n-1}
\label{Eqn: Incomplete beta - B2}
\end{array}
\end{equation}
Substituting the Equations \eqref{Eqn: Incomplete beta - B1} and \eqref{Eqn: Incomplete beta - B2} in the Equation \eqref{Eqn: Incomplete Beta wrt psi}
\begin{equation}
\begin{array}{l}
\displaystyle \frac{\partial}{\partial \psi_{n,m}} \left[ \mathbb{B}(r; a_n, b_n) \right] = \displaystyle c \cdot \int_0^r t^{a_n-1} \cdot (1 - t)^{b_n-1} \left[ \log t - \log(1-t) \right] \ dt
\\[2ex]
\qquad = \displaystyle c \left[ \int_0^r t^{a_n-1} \cdot (1 - t)^{b_n-1} \log t \ dt - \int_0^r t^{a_n-1} \cdot (1 - t)^{b_n-1} \log(1-t) \ dt \right]
\\[2ex]
\qquad = \displaystyle c \left[ \int_0^r \mathbb{B}(a_n, b_n) \cdot f(t; a_n, b_n) \log t \ dt - \int_0^r \mathbb{B}(a_n, b_n) \cdot f(t; a_n, b_n) \log(1-t) \ dt \right],
\\[2ex]
\qquad = \displaystyle c \left[ \int_0^r t^{a_n-1} \cdot (1 - t)^{b_n-1} \log t \ dt - \int_0^r t^{a_n-1} \cdot (1 - t)^{b_n-1} \log(1-t) \ dt \right],
\label{Eqn: Incomplete Beta wrt psi - p2}
\end{array}
\end{equation}
where the last equation is obtained from pdf of beta distribution that is defined as $f(t; a_n, b_n) = \frac{t^{a_n-1} \cdot (1 - t)^{b_n-1}}{\mathbb{B}(a_n, b_n)}$. To solve the above equation, consider the incomplete Beta function $\mathbb{B}(r; a_n, b_n) = \int_0^r t^{a_n-1} \cdot (1 - t)^{b_n-1}$ and differentiate it with respect to the shape parameter $a_n$.
\begin{equation}
\displaystyle \frac{\partial}{\partial a_n}\mathbb{B}(r; a_n, b_n) = \displaystyle \int_0^r \left[\displaystyle \frac{\partial}{\partial a_n} \ t^{a_n-1} \cdot (1 - t)^{b_n-1} \right] \ dt
= \displaystyle \int_0^r \displaystyle \ t^{a_n-1} \cdot (1 - t)^{b_n-1}\cdot \log t \ dt
\label{Eqn: Incomplete derivative wrt a_n}
\end{equation}
Substituting the Equation \eqref{Eqn: Incomplete derivative wrt a_n} in Equation \eqref{Eqn: Incomplete Beta wrt psi - p2},
\begin{equation}
\begin{array}{lcl}
\displaystyle \frac{\partial}{\partial \psi_{n,m}} \left[ \mathbb{B}(r; a_n, b_n) \right] &=& \displaystyle c \left[ \displaystyle \frac{\partial}{\partial a_n}\mathbb{B}(r; a_n, b_n) - \displaystyle \frac{\partial}{\partial b_n}\mathbb{B}(r; a_n, b_n) \right].
\label{Eqn: Incomplete Beta wrt psi - p3}
\end{array}
\end{equation}
Secondly, from Equation \eqref{Eqn: Beta CDF Expansion}, applying $\log$ and differentiating the term $\mathbb{B}(a_n, b_n)$ with respect to $\psi_{n,m}$ on both sides.
\begin{equation}
\begin{array}{rcl}
\displaystyle \frac{\partial}{\partial \psi_{n,m}} \left[ \ln \mathbb{B}(a_n, b_n) \right] &=& \displaystyle\frac{1}{\mathbb{B}(a_n, b_n)} \frac{\partial}{\partial \psi_{n,m}} \left[ \mathbb{B}(a_n, b_n) \right]
\\[2ex]
\displaystyle \frac{\partial}{\partial \psi_{n,m}} \left[ \mathbb{B}(a_n, b_n) \right] &=& \displaystyle \mathbb{B}(a_n, b_n)\cdot\frac{\partial}{\partial \psi_{n,m}} \left[ \ln \mathbb{B}(a_n, b_n) \right]
\end{array}
\end{equation}
Given the definition of Beta function $\mathbb{B}(a_n, b_n) = \frac{\Gamma(a_n)\Gamma(b_n)}{\Gamma(a_n + b_n)}$, where $a_n = \psi_{n,m}c$ and $b_n = (1-\psi_{n,m})c$, the above equation can be written as
\begin{equation}
\begin{array}{l}
\displaystyle \frac{\partial}{\partial \psi_{n,m}} \left[ \mathbb{B}(a_n, b_n) \right] 
\\[2ex]
\quad = \displaystyle \mathbb{B}(a_n, b_n)\cdot\frac{\partial}{\partial \psi_{n,m}} \Big[ \ln \Gamma(\psi_{n,m}\cdot c) + \ln \Gamma((1-\psi_{n,m})\cdot c) - \ln \Gamma(\psi_{n,m}\cdot c + (1-\psi_{n,m})\cdot c) \Big]
\\[2ex]
\quad = \displaystyle \mathbb{B}(a_n, b_n)\cdot\frac{\partial}{\partial \psi_{n,m}} \Big[ \ln \Gamma(\psi_{n,m}c) + \ln \Gamma(c) - \ln \Gamma(\psi_{n,m}c) - \ln \Gamma(\psi_{n,m}c + c -\psi_{n,m}c) \Big]
\\[2ex]
\quad = 0
\label{Eqn: Incomplete Beta wrt psi - p4}
\end{array}
\end{equation}
Finally, substituting the Equations \eqref{Eqn: Incomplete Beta wrt psi - p3} and \eqref{Eqn: Incomplete Beta wrt psi - p4} in the Equation \eqref{Eqn: Beta CDF Expansion},
\begin{equation}
\begin{array}{lcl}
\displaystyle \frac{\partial}{\partial \psi_{n,m}} F(r; a_n, b_n) = \displaystyle c \left[ \displaystyle \frac{\partial}{\partial a_n}\mathbb{B}(r; a_n, b_n) - \displaystyle \frac{\partial}{\partial b_n}\mathbb{B}(r; a_n, b_n) \right].
\label{Eqn: u wrt psi - final}
\end{array}
\end{equation}

\section{More Simulation Results \label{App: More results}}

\begin{figure}[!t]
\floatconts
{Fig: Feedback regret - case1}
{\vspace{-4ex}\caption{Convergence of Feedback Regret with Simulation Experiment}}
{%
\subfigure[Gender Attribute]{%
\label{Fig: Simulation results case1 - gender}
\includegraphics[width=\textwidth]{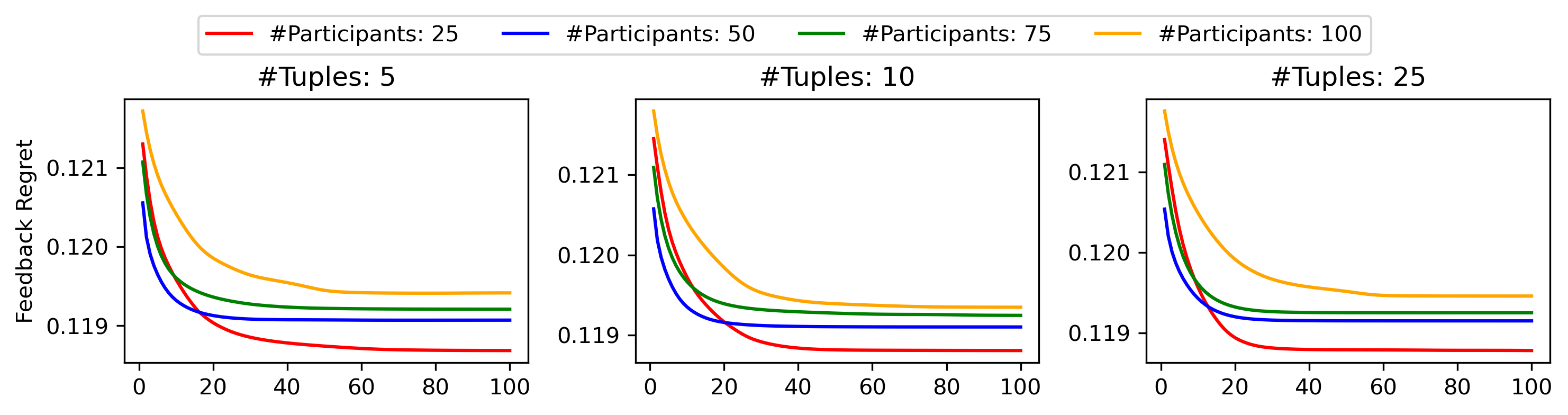}
}\hfill 
\subfigure[Race Attribute]{%
\label{Fig: Simulation results case1 - race}
\includegraphics[width=\textwidth]{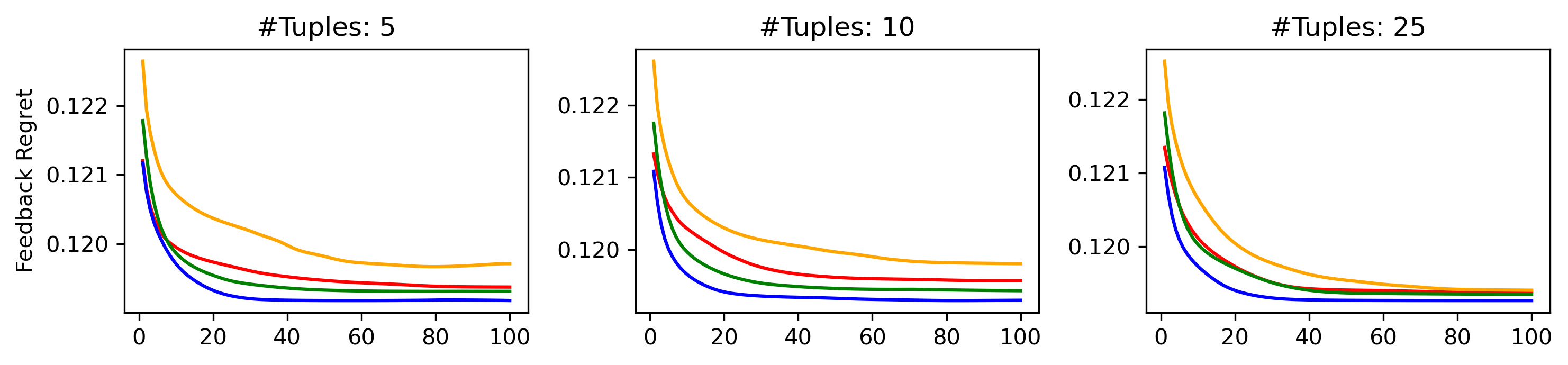}
}\hfill 
\subfigure[Age Attribute]{%
\label{Fig: Simulation results case1 - age}
\includegraphics[width=\textwidth]{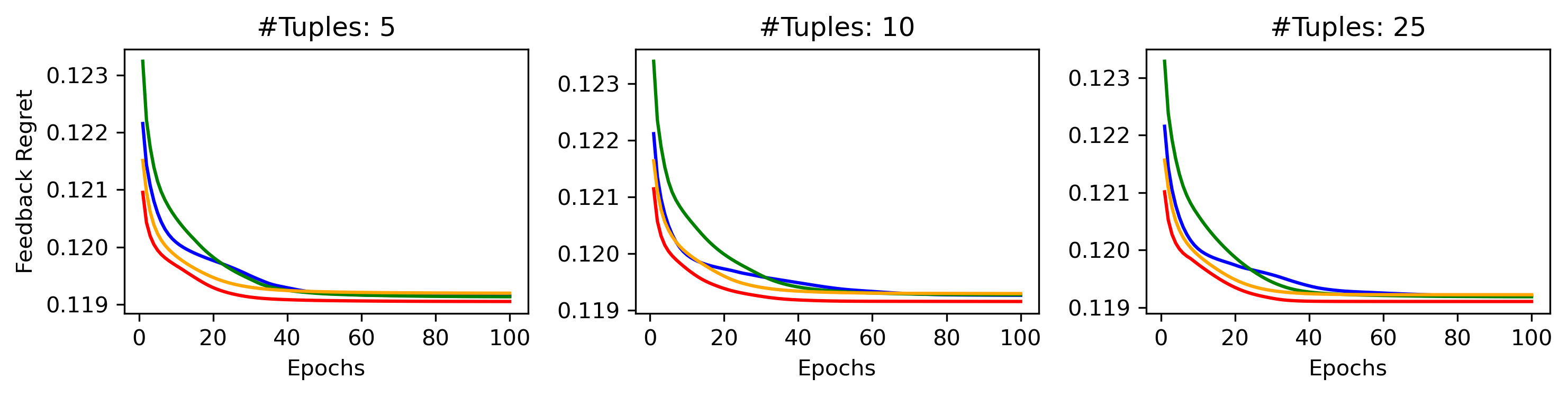}
}
}
\end{figure}

\end{document}